\documentstyle[preprint,prl,aps,epsfig]{revtex}
\newcommand{\be}{\begin{equation}}
\newcommand{\ee}{\end{equation}}
\newcommand{\ba}{\begin{eqnarray}}
\newcommand{\ea}{\end{eqnarray}}
\newcommand{\Var}{{\rm Var}}

\newcommand{\trace}{{\rm trace }}
\newcommand{\bbR}  {{\rm I\!R}}
\newcommand{\pdrv} [2]{\frac{\partial {#1}}{\partial {#2}}}
\newcommand{\vctr} [1]{{\bf #1}}    
\newcommand{\vctrg}[1]{\mbox{\boldmath${#1}$\unboldmath}} 
\newcommand{\mtrx} [1]{{\bf #1}}    
\newcommand{\pL}   [1]{^{[{#1}]}}   
\newcommand{\pLT}  [1]{^{T [{#1}]}} 
\newcommand{\mdl}  [1]{{\tilde {#1}}} 
\newcommand{\spt}  {^{\rm t}}
\newcommand{\spo}  {^{\rm o}}
\newcommand{\spf}  {^{\rm f}}
\newcommand{\spa}  {^{\rm a}}
\newcommand{\spfa} {^{\rm f,a}}
\newcommand{\spfo} {^{\rm f,o}}
\newcommand{\spet} {^{\eta}}
\newcommand{\spep} {^{\epsilon}}
\newcommand{\spG}  {^{{\bf G}}}
\newcommand{\spT}  {^{T}}        
\newcommand{\sbk}  {_{k}}
\newcommand{\sbkn} {_{k-1}}
\newcommand{\mum}  {\mdl{\mu}}
\newcommand{\xh}   {{\hat x}}
\newcommand{\xm}   {\mdl{x}}
\newcommand{\xf}   {x\spf}
\newcommand{\xa}   {x\spa}
\newcommand{\xo}   {x\spo}
\newcommand{\Mkkn} {M_{k,k-1}}
\newcommand{\Bk}   {B\sbk}
\newcommand{\Bm}   {\mdl{B}}
\newcommand{\Bh}   {{\hat B}}
\newcommand{\bb}   {\bar{b}}
\newcommand{\lmdb} {\lambda^{b}}
\newcommand{\bh}   {{\hat b}}
\newcommand{\bm}   {\mdl{b}}
\newcommand{\bk}   {b\sbk}
\newcommand{\bkn}   {b\sbkn}
\newcommand{\ko}   {K\spo}
\newcommand{\KbL} {\bar{K}^{L}}
\newcommand{\Km}   {\mdl{K}}
\newcommand{\vx}   {\vctr{x}}
\newcommand{\vxh}  {\hat{\vx}}
\newcommand{\vxk}  {\vx\sbk}
\newcommand{\vxkn} {\vx\sbkn}
\newcommand{\vy}   {\vctr{y}}
\newcommand{\vyk}  {\vy\sbk}
\newcommand{\vep}  {\vctrg{\epsilon}}
\newcommand{\vet}  {\vctrg{\eta}}
\newcommand{\vomg} {\vctrg{\omega}}
\newcommand{\vomgk}{\vomg\sbk}

\newcommand{\mB}   {\mtrx{B}}
\newcommand{\mBk}  {\mtrx{B}\sbk}
\newcommand{\mBkn} {\mtrx{B}\sbkn}
\newcommand{\mH}   {\mtrx{H}}
\newcommand{\mHk}  {\mtrx{H}\sbk}
\newcommand{\mI}   {\mtrx{I}}
\newcommand{\mK}   {\mtrx{K}}
\newcommand{\mKk}  {\mK\sbk}
\newcommand{\mMkkn}{\mtrx{M}_{k,k-1}}
\newcommand{\mC}   {\mtrx{C}}
\newcommand{\mCk}  {\mtrx{C}\sbk}
\newcommand{\mCkn} {\mtrx{C}\sbkn}
\newcommand{\mG}   {\mtrx{G}}
\newcommand{\mGk}  {\mtrx{G}\sbk}
\newcommand{\mGkn} {\mtrx{G}\sbkn}
\newcommand{\mP}   {\mtrx{P}}

\begin{document}
\begin{center}
\Huge The Kalman-L\'evy filter
\end{center}
\bigskip
\begin{center}
\Large  Didier Sornette$^{\mbox{\ref{ess},\ref{lpec}}}$ and
Kayo Ide$^{\mbox{\ref{das}}}$\\
\end{center}
\begin{center}
Institute of Geophysics and Planetary Physics\\ University of
California Los Angeles at Los Angeles\\ Los Angeles, CA 90095-1567\label{igpp}
\end{center}

\bigskip
\begin{enumerate}

\item Also Department of Earth and Space Sciences, UCLA\label{ess}

\item Also Laboratoire de Physique de la Mati\`ere Condens\'ee, CNRS UMR
6622 and
Universit\'e de Nice-Sophia Antipolis, 06108 Nice Cedex 2,
France\label{lpec}

\item Also Department of Atmospheric Sciences, UCLA\label{das}

\end{enumerate}

\vfill

\begin{flushleft}
Physica D, in press.

{\bf Acknowledgments}: We acknowledge stimulating discussions with
Dan Cayan and Larry Riddle at Scripps and Michael Ghil and Andrew
Robertson at UCLA and thank R. Mantegna for exchanges on how
to generate noise with L\'evy distributions. We thank the two referees
for spotting analytical errors in the submitted version and for
their insightful remarks that helped improve the presentation. All
errors remain ours.
This work was partially supported by ONR N00014-99-1-0020 (KI).
\end{flushleft}

\newpage
\begin{abstract}
The Kalman filter combines forecasts and new observations to obtain an
estimation
which is optimal in the sense of a minimum average quadratic error. The Kalman
filter has two main restrictions: (i) the dynamical system is assumed
linear and
(ii) forecasting errors and observational noises are projected onto Gaussian
distributions. Here, we
offer an important generalization to the case where errors and noises have
heavy
tail distributions such as power laws and L\'evy laws. The main tool needed to
solve this ``Kalman-L\'evy'' filter is the ``tail-covariance'' matrix which
generalizes the covariance matrix in the case where it is mathematically
ill-defined (i.e. for power law tail exponents $\mu \leq 2$). We present the
general solution and discuss its properties on pedagogical examples. The
standard
Kalman-Gaussian filter is recovered for the case $\mu = 2$. The optimal
Kalman-L\'evy filter is found to deviate substantially from the standard
Kalman-Gaussian filter as $\mu$ deviates from $2$. As $\mu$ decreases, novel
observations are assimilated with less and less weight as a small exponent
$\mu$
implies large errors with significant probabilities. In terms of
implementation,
the price-to-pay associated with the presence of heavy tail noise distributions
is that the standard linear formalism valid for the Gaussian case is
transformed
into a nonlinear matrix equation for the Kalman-L\'evy filter. Direct
numerical experiments
in the univariate case confirms our theoretical predictions.

\end{abstract}

\newpage
\tableofcontents

\newpage
\section{Introduction and Motivation}
\label{sec:intro}

The Kalman filter provides the optimal resolution of the problem
of data assimilation under the hypothesis that the system observables
evolve according to linear maps, are linearly related to the true
variables and that the noise acting on the true dynamics and the measurement
errors are mutually uncorrelated and Gaussian.

Two main limitations restrict the performance of the Kalman filter:
\begin{itemize}
\item the nonlinearity of the real system dynamics;
\item the non-normality of the noises.
\end{itemize}
The first item has been addressed partly by using so-called
``extended'' Kalman filters that amount essentially to perform local
linearizations \cite{MGG}.

With respect to the non-normality of the noises,
the general condition for using the Kalman filter is that their covariance
functions exist, which is
satisfied for noise density distributions decaying faster than $1/x^3$.
When the noise distributions are not Gaussian, the validity of the Kalman
filter relies on the existence of a central limit theorem
for state estimators, which exists when the random terms in the model have
arbitrary
distribution with tail decaying faster than $1/x^3$ \cite{Spall}.
For practical applications, the existence of the central limit theorem does not
suffice as a finite observation time may lead to large deviations from the
asymptotic results \cite{Sorbook}. The knowledge of convergence rates
in the central limit theorem are then necessary for the development
of tests of the validity of the model \cite{Aliev}.

Our purpose here is to extend these results to the regime where the existence
of the covariance function is not warranted as occurs for L\'evy
distributions of
noises, or when the covariance functions do exist but
the convergence to the asymptotic result given by the central limit theorem
is extremely slow
making the asymptotic result useless in practice, as
occurs for power law distributions \cite{Sorbook}.
In order to illustrate this idea,
consider the sum of $N$ identically independent distributed random variables,
with a power law probability density function
$1/x^{1+\mu}$ with exponent $\mu>2$. Since the variance
$\sigma^2$
is finite, the central limit theorem applies and the distribution of the sum
converges to the Gaussian law with standard deviation $\approx \sigma 
\sqrt{N}$.
For $N$ finite, the Gaussian law only describes the central part of 
the distribution
of the sum, up to a cross-over $S_0 \approx \sigma \sqrt{N \ln N}$ 
\cite{Sorbook}.
Beyond $S_0$,
the distribution of the sum is a power law with the same exponent $\mu$ and
the weight in probability of this power law tail decays slowly as $\propto
1/(\sqrt{N}~\ln^{3/2} N)$ as $N$ increases. In practice, consider
the three sample sizes $N=10^2, 10^4$ and $10^6$.
The corresponding cross-over values are respectively $S_0 \approx 
2.1$, $3$ and $3.7$
times the standard deviation of the central Gaussian part of the 
distribution of
the sum. These estimates suggest that, even if
the general condition for using the Kalman filter is satisfied
for noise density distributions decaying faster than $1/x^3$,
the Gaussian (or covariance) approach, which is optimal in the
linear least-variance estimation, is not necessarily optimal when
relatively large
fluctuations (of size equal to two standard deviations or larger) occur.

We thus propose to explore how the optimal Kalman filter is modified when the
objective is to minimize, not the variance of the error estimation but, a
natural measure of the large errors, namely the tail of the
distribution of the Euclidean norm of the errors. Our approach thus
develops an alternative class of linear unbiased estimators different from the
standard linear least-variance estimation. Our emphasis is in trying
to control and minimize the large (and rare) errors,
with the penalty that the variance, if it exists, will
be sub-optimal compared to the standard linear least-variance estimation.
In this goal, we focus our attention on noises in the observable and the
unobservable variables both distributed with a power law tail
of fixed exponent $\mu$ and known amplitudes (scale factors). This
assumption offers a well-defined theoretical limit of ``large fluctuations'',
putting the emphasis on the complement to the ``small noise'' limit
captured by the standard linear least-variance approach.

In addition to its pure theoretical interest, we observe that
many systems in Nature are claimed to exhibit power law
distributions \cite{Sorbook}
and the present results may thus have direct application.
Power law distributions have been found to quantify the size-frequency
Gutenberg-Richter distribution of earthquakes, of
  hurricanes \cite{Barton}, of volcanic eruptions, of floods,
of meteorite sizes and so on. The distribution of seismic
fault lengths is also documented to be a
power law with exponent $\mu \simeq 1$. In the insurance business,
recent studies has shown that the
distribution of losses due to business interruption
resulting from accidents \cite{Zaj1,Zajx} is also
a power law with $\mu \simeq 1$.

Several previous works have addressed related issues.
Le Breton and Musiela's work \cite{Lebreton} is closest to ours but has
limitations: (i) it is based on a continuous time description of the
dynamical and observation processes and is thus more difficult to
apply to concrete situations ; (ii) it minimizes the difference
between the
true dynamics and a filtered observation in the $L^{\mu}$-norm sense where
$\mu$
is the index exponent of the L\'evy distributions; in other words
it relies on an explicit solution of the dynamics; in this way, Le Breton
and Musiela
circumvent the two delicate questions of generalizing the covariance matrix
and of choosing
the objective function to optimize; (iii) it does not really address
the genuine Kalman problem which consists in {\it mixing} forecasts and
observations.
Recently, Ahn and Feldman \cite{Ahn} have proposed a filter
for the case where the signal is Gaussian while the observation noise is
a L\'evy process. Their filter is optimal in the sense of
minimizing the $L^2$ error, i.e. the distance between the true
dynamics and the filter output.
This choice of the $L^2$ is realizable in their case
because the signal is assumed Gaussian and the
integrability assumption is thus satisfied. The problem however is that the
optimal filter
will depend in general upon the choice of the norm that measures the
prediction error.
In addition, due to the rather intractable nonlinear recursive solution
they obtain,
they propose a sub-optimal filter for numerical purpose.

Our approach circumvents these difficulties by focusing on large errors.
Our goal is thus to minimize the large errors between the analysis
(obtained by assimilation of observation with prediction) and the
true trajectory. With this choice, the technical problem we have to solve
is to characterize the tails of error distributions, which are a function of
the tail of the distributions of individual dynamical variables and of noises
and of their mutual dependence. In order to determine these 
dependences, we propose to
use the concept of a ``tail-covariance'' matrix
that generalizes the standard covariance matrix of the analyzed signal to
the case of power laws and
L\'evy distributions. Tail-covariance matrices are constructed
as the matrices of {\it signed} scale factors (the scale
factors being defined as the global amplitudes of the
power law tails)
of the distribution of all
the products of the system and forecasted variables. The main idea is thus
to replace
the characterization of errors and of correlations by the ensemble of
distributions
of the products of all possible pairs of variables.
Our approach thus replaces a reasoning based on the second centered moments
into
one based on the tails of the distribution of deviations from true values.
It is shown
that the natural error amplitude to be minimized is the trace of the
tail-covariance,
which is a straightforward extension of
the standard approach which determines the Kalman filter by minimizing the
average errors quantified by the trace of the covariance matrix of the
analyzed signal.

The paper is constructed in a pedagogical manner, from the simple
univariate filter problem
to the full multivariate Kalman filter. In section 2, we formulate the
problem for the
univariate filter with power law noises, corresponding to the situation
where only novel
observations are assimilated without mixing with a dynamical forecast. A
detailed
discussion is offered to compare the power law case with the standard
Gaussian case.
In section 3, we generalize this problem to the case of multivariate
estimators.
In section 4, we address the Kalman problem of mixing both forecasts and
observations with
power law and L\'evy law errors and develop our general solution. The
special univariate
case is studied in great details by numerical experiments to contrast 
the performances of the
Kalman-L\'evy
with those of the standard Kalman-Gaussian filter.

\section{Univariate estimator: Principle of Estimation with Power Laws}
\label{sec:uni}%
\subsection{Formulation of the Problem}
\label{sec:uni-problem}

Take two samples $\xf$ and $\xo$ of an observable state variable $x$;
the superscripts $\{\cdot\}\spfo$ correspond to forecast and
observation of the data assimilation system, respectively.
The two samples are contaminated by independent noise $\omega\spf$
and $\omega\spo$.
The estimate $\xh$ of $x$ is sought as a linear combination
of $\xo$ and $\xf$ with the corresponding {\it positive} weights $K\spf$
and $\ko$
\be
\xh = K\spf \xf + \ko \xo~.  \label{eq1}
\ee
One of our main goals is to determine the optimal weights so as to minimize
the resulting uncertainty of $\xh$.
We make two assumptions concerning samples in this study.

The first assumption that the two samples are unbiased
\be \label{eq:unbias}
\langle \xf \rangle = \langle \xo \rangle = \langle x \rangle~,
\ee
where $\langle z \rangle$ is the expectation of $z$.
Expectation of the state variable $\langle x \rangle$ is not known a priori.
By requiring further that the estimate should be unbiased
$\langle\xh\rangle=\langle x \rangle$, we obtain a relation between
the two weights $K\spf+\ko=1$ which allows us to rewrite (\ref{eq1}) as
\be
\xh = \xf + \ko (\xo - \xf) ~,   \label{eqsd}
\ee
and therefore reduces the problem to the determination of a single unknown
$0 \leq \ko \leq 1$.
This expression (\ref{eqsd}) can be interpreted as a filtering
of the observed data $\xo$ into the dynamical forecast $\xf$ by a
weighted increment $\ko (\xo - \xf)$.

The second assumption is that both sample errors,
\be
\omega\spfo = x\spfo - x~,  \label{eq:omgfo}
\ee
are distributed according to a power law as defined in (\ref{jnjkaka}),
or according to a L\'{e}vy law
as defined in Appendix \ref{Appendix:levy}, with the same exponent
$\mu\spfo\equiv\mu$.
The important property for our purpose is that the tails of the
probability density functions of the two sample errors
are independent and given by
\be
P(\omega\spfo) \simeq {C\spfo_{\pm} \over |\omega\spfo|^{1+\mu}}
  ~~~~~~~\mbox{  $|\omega\spfo| \longrightarrow \pm \infty$}~,
\label{jnjkaka}
\ee
where the subscript $\{\cdot\}_{\pm}$ reflects that the tail
distribution can be asymmetric depending on the sign of $\omega\spfo$.
In this study, we focus on the symmetric case, i.e.,
$C\spfo_{\pm}\equiv C\spfo$.

The family of power laws is
characterized by two parameters, the exponent $\mu$ and the
`scale factor' $C$.
The exponent $\mu$, on one hand, controls the decay rate of the
probability as well as its scaling (or self-similar) properties.
The scale factor $C$, on the other hand, controls the overall
amplitude of the power law tail,
i.e., the larger it is, the more important is the power law tail.
More precisely, if the power law tail (\ref{jnjkaka}) holds for
$\omega$ larger than some minimum value $\omega_{\rm min}\spfo$, the
weight in probability of the power law, i.e. the probability that
$\omega$ is larger than $\omega_{\rm min}\spfo$ is
$(C\spfo/\mu)(\omega_{\rm min}\spfo)^{-\mu}$.
As shown in Appendix \ref{Appendix:levy} in the case of L\'{e}vy laws, the
scale parameter fully characterizes the distribution for all
variations (and not only in the tail).

Notice that (\ref{jnjkaka}) can be rewritten in terms of the
dimensionless variable  $\omega/C^{1 \over \mu}$ with the superscripts
dropped for simplicity
\be
P(\omega) d \omega \simeq
{1\over |\omega/C^{1 \over \mu} |^{1+\mu}}
   d\left(\omega/C^{1 \over \mu}\right)
~~~~~
\mbox{  $|\omega| \longrightarrow \pm \infty$}~,
\label{jnadsfajkaka}
\ee
showing that  $C^{1 \over \mu}$ is the characteristic scale of the
self-similar fluctuations of $\omega$.
For $\mu \leq 2$ (resp. $\mu \leq 1$), the variance (resp. mean) is
not defined mathematically.
We shall see the effect of these $\mu$-dependent properties throughout this
study.

Using the continuation property given in \cite{Feller}
(Problem 15 of Section 12) on the characteristic
function of distributions with power tails, we obtain the
following results. Let us call a $\mu-$variable a variable with
a distribution function with a power law tail. Then,
\begin{enumerate}
\item[(i)]  if $w_i$ and $w_j$ are two independent $\mu$-variables
characterized by the scale factors $C_i^\pm$ and $C_j^\pm$, then
$w_i + w_j$ is also a $\mu$-variable with $C^\pm$ given by
$C_i^\pm + C_j^\pm$.

\item[(ii)]  If  $w$ is a $\mu$-variable with scale factor $C$, then 
$p \times w$
(where $p$ is a real number) is a $\mu$-variable with scale factor
$p^\mu C$. If $p<0$ and the distribution of $w$ is symmetric, then
$p \times w$ is a $\mu$-variable with scale factor
$|p|^\mu C$.

\item[(iii)] If $w$ is a $\mu$-variable, then ${\rm sign}(w) |w|^q$, 
with $q>0$, is a
$\mu \over q$-variable.

\item[(iv)] If $w_i$ and $w_j$ are two independent $\mu$-variables, then
the product $x={w_iw_j}$ is also a $\mu$-variable up to logarithmic
corrections.
\end{enumerate}

Using the rules (i) and (ii), we find that the distribution of $\xh$ is
also a power law (\ref{jnjkaka}) with the same exponent $\mu$,
but with an adjusted scale factor ${\hat C}$ given by
\be
{\hat C} = (1-\ko)^{\mu} C\spf + (\ko)^{\mu} C\spo~.
  \label{faohgal}
\ee

The expression (\ref{faohgal}), which is valid for
$0 \leq \ko \leq 1$ such that the scale factors remain positive,
can be reduced to the usual result for the Gaussian distributions
\cite{Malanotte}
\be
{\hat \sigma}^2 = (1-\ko)^{2} (\sigma\spf)^2 +
                    (\ko)^{2}  (\sigma\spo)^2~,
\label{fdhglahgajnv}
\ee
by setting the exponent $\mu=2$ and replacing the
scale factors $C\spfo$ with $(\sigma\spfo)^{2}$.
We thus see that the scale parameter $C$ is the generalization of the
variance $\sigma^2$. The technical reason for this comes
>from the form of the characteristic function of a distribution with
a power law tail, as given in \cite{Feller}
(Problem 15 of Section 12). The situation is even simpler to discuss when
considering symmetric L\'evy laws whose
characteristic functions read
\be
\hat{L}_{\mu}(k)=\exp\left(-a_{\mu}|k|^{\mu}\right)~,~~~~{\rm for}~0 
< \mu \leq 2~,
\label{bbvvvcnc}
\ee
where $a_{\mu}$ is a constant proportional to the scale parameter
$C$ \cite{GK}. The important point is that, for $\mu$ strickly less than $2$,
the inverse Fourier transform of $\hat{L}_{\mu}(k)$ gives a power law tail,
while for $\mu=2$, it gives a Gaussian law. The continuity between
the expressions (\ref{faohgal}) and (\ref{fdhglahgajnv}) can thus be 
traced back
to that of $\hat{L}_{\mu}(k)$ as a function of $\mu$ at $\mu=2$.
(see Appendix \ref{Appendix:levy}
for a more formal derivation of this fact).

\subsection{Solution}
\label{sec:uni-solution}

The standard optimal estimation methodology consists in
minimizing the variance ${\hat \sigma}^2$ with respect to the weight
factor $\ko$.
The solution for $\ko$ then gives the best weighting in the
sense that we remain with the smallest uncertainty from
the estimation with the novel data, by minimizing the expectation
distance between $\xh$ and $x$ in the mean-square sense.
In order to generalize this methodology to the situation where the
errors are distributed according to power law distributions, we
propose the following central idea, i.e., to minimize the scale factor
${\hat C}$ with respect to $\ko$
\ba
\pdrv{}{\ko}{\hat C}&=&
  \mu\left[-(1-\ko)^{\mu-1}C\spf+(\ko)^{\mu-1}C\spo\right]=0
  \label{jfjks}  \\
\pdrv{^{2}}{(\ko)^{2}}{\hat C}&=&
  \mu(\mu-1)\left[(1-\ko)^{\mu-2}C\spf+(\ko)^{\mu-2}C\spo\right]\geq 0.
  \label{eq:optko}
\ea
The justification for this procedure is that the uncertainty in the
estimation $\xh$ of $x$ is inescapably distributed according to a
power law distribution with the same exponent $\mu$
as a result of the rules (i)-(ii) given above.
Consequently, the optimization using the weight $K^{o}$ can
be performed only in one purpose, namely to decrease the global
amplitude of the  power law controlled by ${\hat C}$ but without be able to
distort its shape defined by $\mu$.
The optimal weight $\ko$ as the solution to (\ref{jfjks}) depends
on the value of the exponent $\mu$ as follows.
\begin{enumerate}
\item For $\mu > 1$, the minimization of ${\hat C}$ given by
(\ref{faohgal}) with respect to $\ko$ gives
\be
\ko = {1 \over 1 + \lambda^{\mu \over \mu-1}}~,
\label{qqwersfqt}
\ee
where
\be
\lambda \equiv {(C\spo)^{1\over \mu} \over (C\spf)^{1\over \mu}}
\label{fbbrfjkfwq}
\ee
is the ratio of the characteristic error size of the two
samples, as defined in the distribution (\ref{jnadsfajkaka}).
The resulting optimal scale factor is
\be
C\spa = {\lambda^{\mu} \over
    \left(1 + \lambda^{\mu \over \mu-1}\right)^{\mu-1}}~ C\spf~
  \label{eq:CaL}
\ee
where the superscript $\{\cdot\}\spa$ stands for analysis according
to the data assimilation convention.

\item For $\mu<1$, there is no optimal solution for $\ko$ which is not on
the boundaries of the search interval and that minimizes
${\hat C}$, because $\partial^{2}{\hat C}/\partial(\ko)^{2}<0$ violates
the second condition in (\ref{eq:optko}).
Physically this implies that the fluctuations are so wild that
the estimation by the weighted average is not a
good strategy, and that only the measurement with the smallest scale
factor should  be kept for the estimation of $\xh$
\be
\ko = \left\{ \begin{array}{lll}
0 & \qquad & \mbox{for $C\spo> C\spf$} \\
1 & \qquad & \mbox{for $C\spo< C\spf$} \end{array}
	\right.~, \label{fhafaa}
\ee
and therefore
\be
C\spa = \left\{ \begin{array}{lll}
  C\spf & \qquad & \mbox{for $C\spo> C\spf$} \\
  C\spo & \qquad & \mbox{for $C\spo< C\spf$} \end{array}
	\right.~. \label{eq:CaLheavy}
\ee
This second case $\ko = 1$, consisting in
trusting observations over the forecast, is known as ``direct substitution''
\cite{Daley}. Here, we have shown that it constitutes indeed the best
strategy in the specific case $\mu \leq 1$ and $C\spo< C\spf$.
Note that a similar solution applies when the two observations
have different exponent $\mu$: full weight $\ko =0$ or $1$ should be put on the
observation with the largest exponent, as it has the smallest fluctuations.
This solves the general case as well.

\end{enumerate}

To make the problem interesting,
in this paper we consider all noise sources to have the same exponent
$\mu$, so that
the problem is a ``fight between scale-factors''. This case is not as
restricted
as it would appear at first site: if the mechanisms leading to the power
law tail are
intertwinned, such as for instance with a common source of
underlying multiplicative noise,
it can be shown \cite{Kesten,goldie}
that the power law exponent of the different variables will
be the same
as soon as there is non-vanishing coupling between the variables.
This case will be investigated in a forthcoming work.

The following argument retrieves (\ref{fhafaa}).
When neither the variance nor the mean exist, and when the minimization
(\ref{jfjks}) of the
scale factor becomes meaningless, the last natural quantity to estimate is
the probability that the error remaining
after assimilation is smaller than the error on the two measurements.
Suppose that we have the knowledge that the errors in the second
measurement are larger in
probability than that of the first measurement, i.e. $C\spo> C\spf$.
We then require the maximization of the probability $P_{\rm improvement}$ that
\be
|(1-\ko) \omega\spf + \ko \omega\spo| \leq |\omega\spf|~.
\label{fjnafafdvgre}
\ee
Let us assume that $\omega\spf$ is found positive.
Then, this probability is the same as the probability that
\be
\left(-{2 \over \ko} + 1 \right) \omega\spf
   \leq \omega\spo \leq \omega\spf~.
\ee
The probability for (\ref{fjnafafdvgre}) to be verified is thus
\be
P_{\rm improvement} = 2 \int_0^{\infty} d\omega\spf P\spf(\omega\spf)
   \int_{\left(-{2 \over \ko}+1\right) \omega\spf}^{\omega\spf}
      P\spo(\omega\spo) d\omega\spo~,   \label{fbbgaljvqa}
\ee
where the factor $2$ comes from the counting of the cases where
$\omega\spf$ can be found negative.
By taking the derivative of (\ref{fbbgaljvqa}) with respect to
$\ko$, we obtain
\be
{d P_{\rm improvement} \over d \ko} =
-{4 \over (\ko)^2}
  \int_0^{\infty} d\omega\spf ~ \omega\spf ~ P\spf(\omega\spf)
  P\spo\left(\left(-{2 \over \ko} + 1 \right) \omega\spf\right) ~,
\label{fzzzsajvqa}
\ee
which is always negative.
Thus, the probability that the error
is reduced is maximum for $\ko = 0$, i.e. without assimilating the new
observation.
Intuitively, the power law tails with exponent $\mu<1$ are so ``wild'' that
it is preferable to keep only the observation with the smallest scale factor.
A similar derivation holds
in the case where the errors in the second measurement are smaller in
probability than that of the first measurement, i.e. $C\spo < C\spf$:
the probability that the error
is reduced is maximum for $\ko = 1$, i.e. with the assimilation of the new
observation
and the rejection of the first one.
Again, the observation with the smaller scale factor is preferred and the
other is
rejected in this ``wild tail'' regime $\mu \leq 1$.

\subsection{Properties of the ``L\'evy-estimator'' solution}
\label{sec:uni-prop}

\begin{figure}
\begin{center}
\epsfig{file=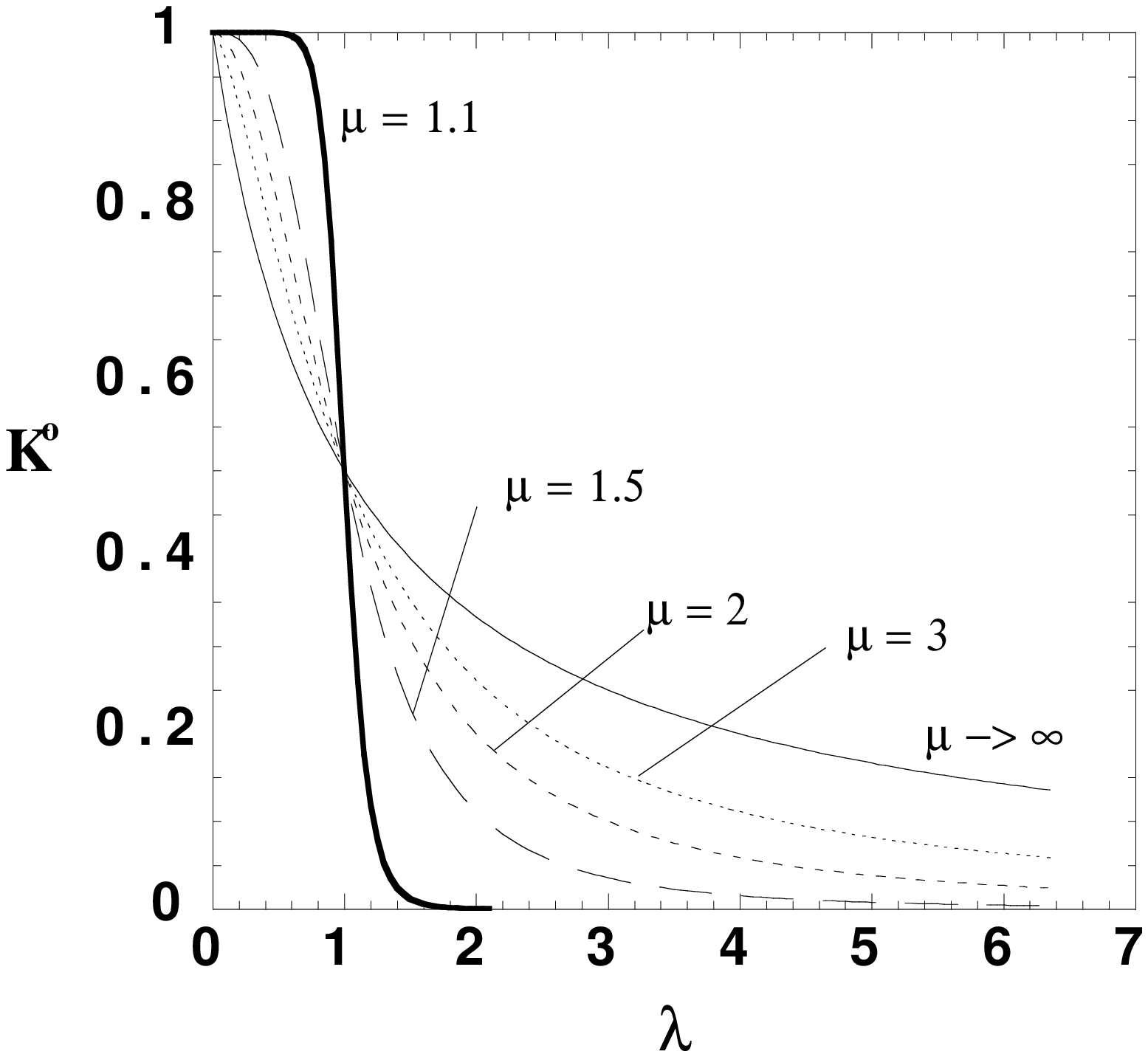,height=10cm}
\caption{\protect\label{figmu} Dependence of the weight $\ko$
given by (\ref{qqwersfqt}) to the second measurement as a function
of the relative amplitude $\lambda$ (eq.(\ref{fbbrfjkfwq}) of the 
noise of the two
measurements:
a small (resp. large) value of  $\lambda$ corresponds to a small
(resp. large) error on the second measurement relative to the first one.
The different curves correspond to different tail exponents $\mu$.}
\end{center}
\end{figure}

We now examine the fundamental properties of the optimal weight
$\ko$ given by  (\ref{qqwersfqt}) which holds when the
distributions of errors are pure L\'{e}vy laws as well as when they
only exhibit a power law tail controlling the large variations.
Figure \ref{figmu} shows the influence of the tail exponent $\mu$ on
the optimal weight $\ko$ as a function of the error ratio $\lambda$
of the two measurements given by (\ref{fbbrfjkfwq}).
As $\mu$ approaches $1$, $\ko$ crosses over very sharply from
one to zero when $\lambda$ goes through $1$, recovering the
regime $\mu \leq 1$ given by (\ref{fhafaa}).
For larger $\mu$'s, the transition of $\ko$ from $1$ to $0$ is
smoother as $\lambda$ varies.

The result (\ref{qqwersfqt}) for $\mu=2$ holds not only for the power
law distribution itself with
$\lambda=(C\spo)^{1\over 2}/(C\spf)^{1\over 2}$ but also for the
Gaussian law distribution with the weight
\be
K^{G}={1 \over 1 + (\lambda^{G})^2}~,
\label{qqgarwerqat}
\ee
where the superscript $\{\cdot\}^{G}$ corresponds to the Gaussian,
and
\be \label{eq:lambdaG}
\lambda^{G}\equiv{\sigma\spo\over \sigma\spf}~
\ee
is the ratio of the characteristic error size according to the
Gaussian law.
Such $K^{G}$ minimizes the variance ${\hat \sigma^{2}}$
given by (\ref{fdhglahgajnv}).
This is natural since the stable
L\'{e}vy law with $\mu = 2$ is nothing but the Gaussian law
with the exact correspondence $C\spfo=(\sigma\spfo)^2$ (see Appendix 
\ref{Appendix:levy}:
there should be no confusion between the exponents $\mu$ defining the
characteristic functions of stable laws and the exponents $\mu$ of
arbitrary power laws).
Thus, the curve for $\mu=2$ in Figure \ref{figmu} also applies
for the Gaussian law with $\ko=K^{G}$ and
$\lambda=\lambda^{G}$.
The result (\ref{qqwersfqt}) reflects the impact of the relative
uncertainties in
$\xf$ and $\xo$ that are quantified by a parameter depending on the
ratio of characteristic error size
$\lambda^{{\mu\over\mu-1}}=\left(C\spo/C\spf\right)^{1\over\mu-1}$.

The weight $\ko$ also represents the normalized increment $(\xh-\xf)$
added to the initial difference $(\xo-\xf)$ as seen from (\ref{eqsd}):
\ba
  \ko={\xh-\xf \over \xo-\xf}~.
\ea
Figure \ref{figmu} therefore can be interpreted as showing the normalized
increment depending on the tail exponent $\mu$, with the extreme
cases $\xh=\xo$ at $\ko=1$ and $\xh=\xf$ at $\ko=0$.
At $\lambda=1$ where $\xf$ and $\xo$ has the same uncertainty
in terms of scale factor $C\spf=C\spo$ (\ref{fbbrfjkfwq}),
$\ko=0.5$ puts $\xh$ at the exact center point between $\xf$ and
$\xo$ for any $\mu$.
For $\lambda<1$ (resp.$\lambda>1$) where $\xo$ with scale factor $C\spo$ is
more (resp. less) accurate than $\xf$ with scale factor $C\spf$,
the smaller the exponent $\mu$ is, the closer $\xh$ is to the more
accurate sample $\xo$ (resp. $\xf$).
The estimation by the weight $\ko$ for the heavier tail
distributions with $\mu < 2$ therefore favors the accurate sample more
strongly than in the least-variance case.
Interestingly, this situation is reversed for power law tails with
exponents $\mu > 2$. i.e. the weight favors the accurate sample less
strongly than in the Gaussian case.
This situation applies in particular to exponential distributions that are
formally
obtained as the limit $\mu \to \infty$.

\subsection{Quality of Improvements: L\'{e}vy versus Gaussian estimators}

\subsubsection{Case $\mu >2$}

Let us investigate the pros and cons of the solution (\ref{qqwersfqt})  for
$\ko$
as the  optimal weight for power law tails
in contrast to its Gaussian counterpart (\ref{qqgarwerqat}) giving $K^{G}$.
In this goal, we propose a specific example using the Student's
distribution with $\mu$ degrees of freedom, whose density function
\cite{Johnskotz},
\be
P_{\mu}(\omega) = {\Gamma\left({\mu+1 \over 2}\right) \over
  \sqrt{\mu \pi}~\Gamma\left({\mu \over 2}\right)}~
  {1/s \over
   \left[1+\left({\omega \over s \sqrt{\mu}}\right)^2\right]
    ^{1+\mu\over 2}}~, \label{fdanhfal}
\ee
is defined for $-\infty < \omega < +\infty$.
The Student's distribution $P_{\mu}(\omega)$ has a bell-like
shape like the Gaussian (and actually tends to the Gaussian
in the limit $\mu \to \infty$).
It is however a power law like (\ref{jnjkaka}) for large
$|\omega|$ with a tail exponent equal to the number
$\mu$ of degrees of freedom defining the Student's distribution,
with a scale factor
\be \label{eq:Cmu}
C_{\mu}(s) = {\Gamma\left({\mu +1 \over 2}\right) \over \sqrt{\mu \pi}
~\Gamma\left({\mu \over 2}\right)} ~ \mu^{1+\mu \over 2}~
s^{\mu}~.
\ee
The parameter $s$ represents the typical width of the Student's
distribution.
The variance exists only for $\mu>2$ and is given by
\be
\Var \equiv \sigma^2 = {\mu \over \mu -2}~ s^2~.
\label{fblvbaa}
\ee

We assume that the forecast (resp. observation)
sample $\xf$ (resp. $\xo$) has an error $\omega\spf$ (resp. $\omega\spo$)
distributed according to the student's distribution (\ref{fdanhfal})
with typical  width $s\spf$ (resp $s\spo$) but with the same exponent
$\mu$.
The L\'{e}vy weight $K^L$ given by (\ref{qqwersfqt}) and the standard
Gaussian weight $K^{G}$ given by (\ref{qqgarwerqat}) are represented
by the same error ratio
\be \label{eq:lambdas}
\lambda^G=\lambda^{L}={s\spo \over s\spf} = \lambda~.
\ee
It is worth recalling that $\ko$, which we denote here $K^L$,
  given by (\ref{qqwersfqt}) is obtained
so as to minimize the scale factor $C_{\xh}$ given by (\ref{faohgal}),
while $K^{G}$ given by (\ref{qqgarwerqat})
minimizes the  variance $\Var_{\xh}$ expressed by
(\ref{fdhglahgajnv}).
The impact of the difference between these two weights can be
quantified in several ways for $\mu>2$ where the variance
exists.

\begin{figure}
\begin{center}
\epsfig{file=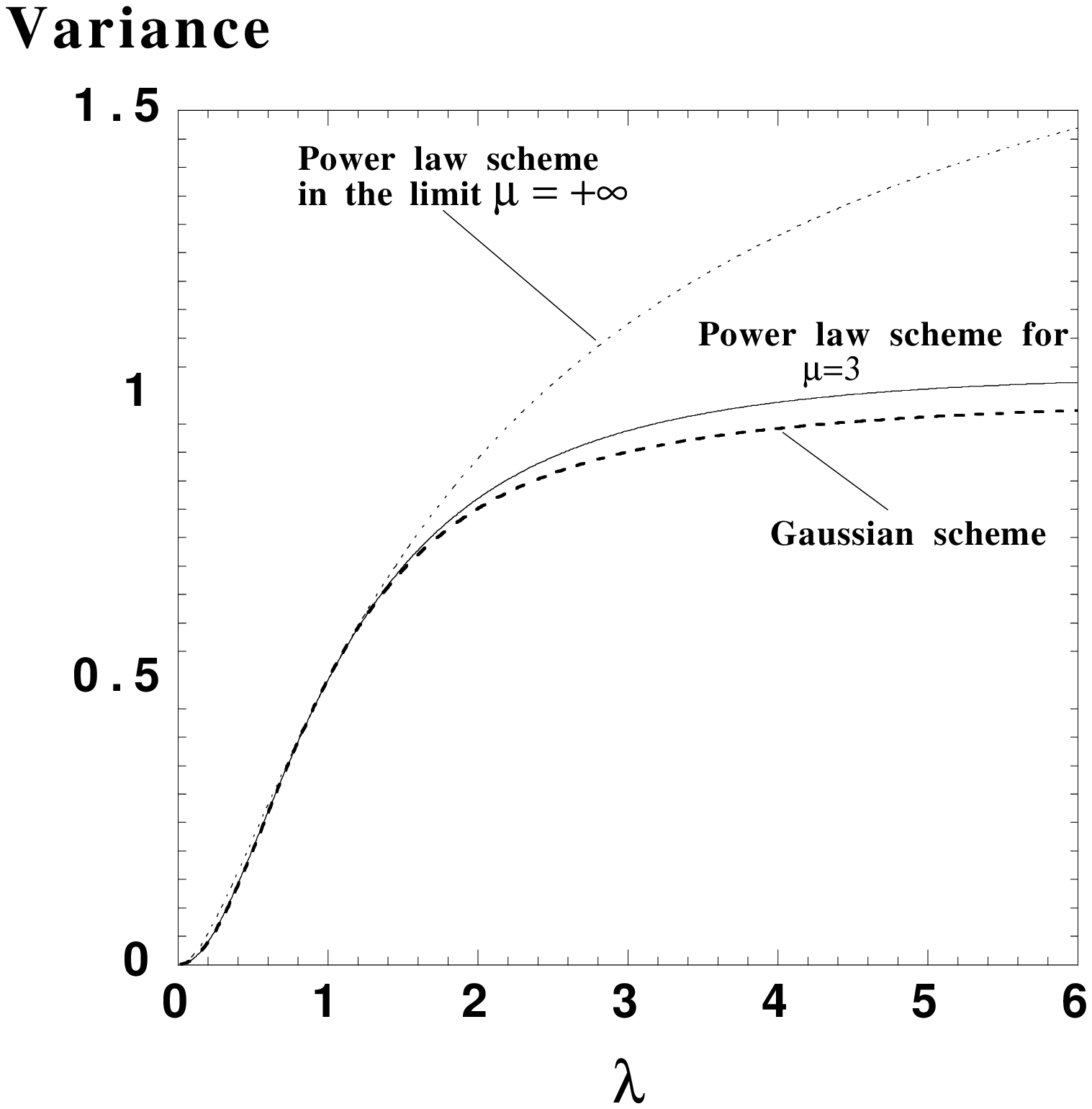,height=10cm}
\caption{\protect\label{varlambd} Dependence of the variance
$\Var_{\xh}^{L}$ and $\Var_{\xh}^{G}$ of the total
error obtained using respectively the L\'{e}vy and the
Gaussian weights, as a function of $\lambda = {s\spo \over s\spf}$
equal to the ratio of the typical widths of the Student's distributions for the
two measurements. }
\end{center}
\end{figure}

One measure is the corresponding variance
$\Var_{\xh}=(1- \ko)^2(\sigma\spf)^2
                    +(\ko)^2(\sigma\spo)^2$
of the total error $(1-\ko) \omega\spf + \ko \omega\spo$
given by
\ba \label{eq:VarL}
\Var_{\xh}^{L} &=&
  {\left(1+\lambda^{2\over\mu-1}\right) \lambda^{2} \over
   \left(1+\lambda^{\mu\over\mu-1}\right)^{2}}(\sigma\spf)^2 \\
\Var_{\xh}^{G} &=& {\lambda^2 \over 1+\lambda^{2}}~
(\sigma\spf)^2~, \label{eq:VarG}
\ea
where $\Var_{\xh}^{L}$ (resp. $\Var_{\xh}^{G}$) is the variance obtained by
using
the solution $K^L$ (resp. $K^G$).
Figure \ref{varlambd} shows $\Var_{\xh}^{L}$ and
$\Var_{\xh}^{G}$ as a function of $\lambda$ for $\mu=3$:
by construction, we verify that the variance of the total
error is less with $K^{G}$ than with $\ko$. This is expected since, by
construction,
$K^{G}$ minimizes the variance. However, the difference is small, less than
$10\%$.
Anyway, this measure would then
suggest that the Gaussian filter is better.

\begin{figure}
\begin{center}
\epsfig{file=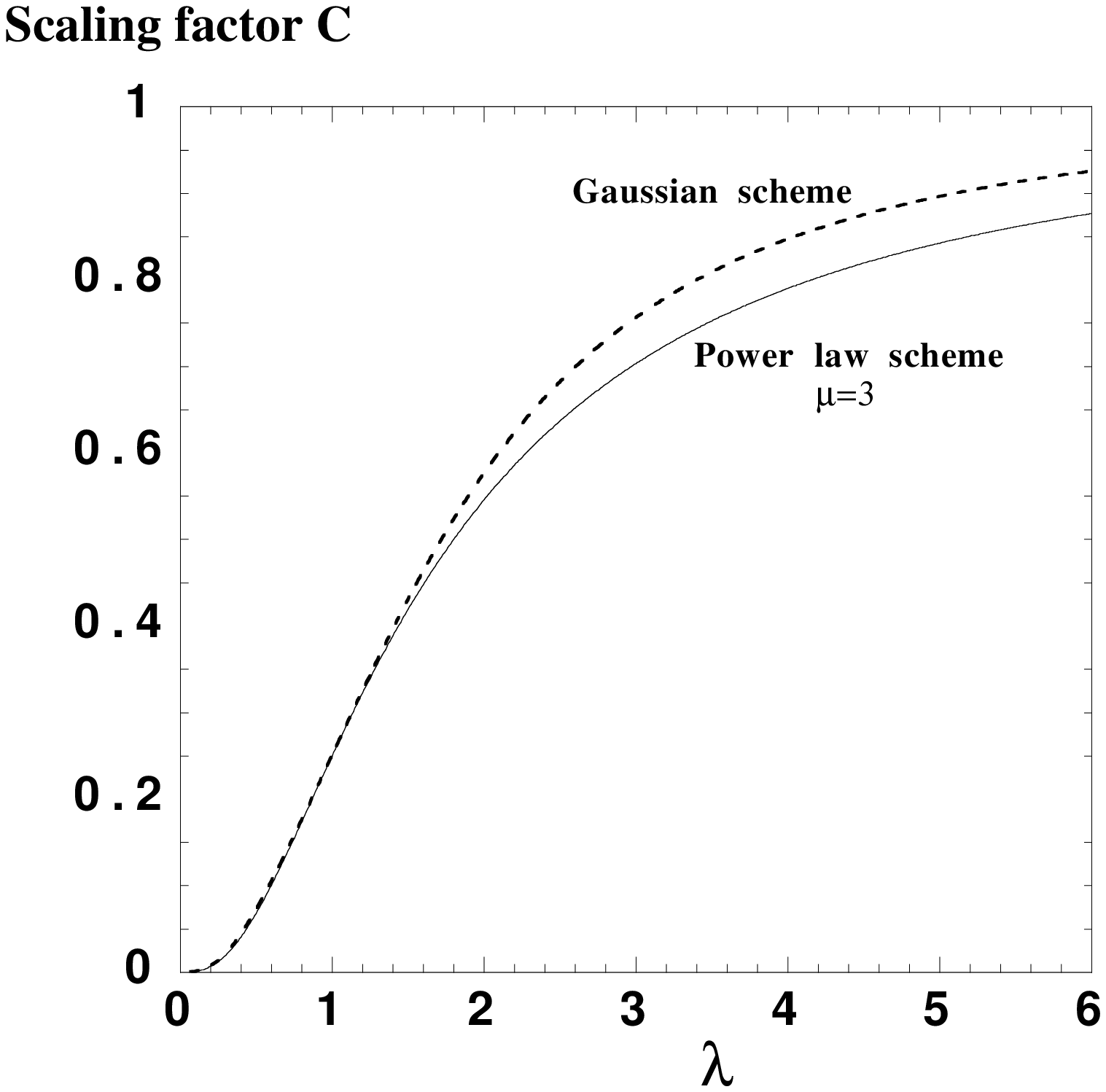,height=10cm}
\caption{\protect\label{scalefaclambd} Dependence of the scale factors
$C_{\xh}^{L}$ and $C_{\xh}^{G}$ of the total error obtained
using respectively $\ko$  and $K^{G}$ weights,
as a function of $\lambda$. }
\end{center}
\end{figure}

However, for power law distributions of errors, the variance is
well-known to be a rather poor representation of the variability,
especially in the tail.
It is thus interesting to compare the scale factors $C_{\xh}^{L}$
and $C_{\xh}^{G}$ obtained in the two schemes since they quantify
the total weight of the power law tails.
We determine the scale factors for the L\'{e}vy and Gaussian weights as
another measure of the goodness of the filtering method:
\ba
C_{\xh}^{L} &=&
  {\lambda^{\mu} \over
    \left(1 + \lambda^{\mu \over \mu-1}\right)^{\mu-1}} C\spf~,
\label{jnbvchhd}\\
C_{\xh}^{G} &=& {\lambda^{\mu} \over
   \left(1 + \lambda^{2}\right)^{\mu} (1 + \lambda^{\mu})^{-1}} C\spf~.
  \label{eq:CG}
\ea
$C_{\xh}^{L}$ (resp. $C_{\xh}^{G}$) is obtained by putting $K^L$ (resp. $K^G$)
in expression (\ref{faohgal}).
Figure \ref{scalefaclambd} shows the scale factors $C_{\xh}^{L}$ and
$C_{\xh}^{G}$ as a function of $\lambda$ for $\mu=3$:
the weight $K^L$ is now found to be better than the usual
Gaussian weight $K^{G}$, since a smaller scale
factor implies
smaller probabilities for large fluctuations. The improvement is however not
very large, typically of the order of or less than $10\%$, i.e. of the same
order as the difference between the variances (but in reverse ranking).
These relatively small differences between
the Gaussian and L\'evy filtering procedures become enormous for
the case $\mu < 2$ discussed next.

The comparison between Figures \ref{varlambd} and \ref{scalefaclambd}
shows that one cannot achieve simultaneously the minimization of the
variance of the error and the minimization of the weight of the tail
of large deviations of the error:  either one or the other can be optimized.

\subsubsection{Case $\mu <2$}

The situation is dramatically different when $\mu < 2$, for which the
variance is not mathematically defined.
In this case, an empirical determination of the variance is very
unstable and absolutely unreliable.
The standard Gaussian weight $K^{G}$ is completely useless.
In contrast, the L\'evy weight $K^L$ gives a simple and clear-cut
recipe that allows one to optimize large fluctuations in the
weighting procedure.

Let us illustrate this result by the following numerical experiments
using the Cauchy distribution for the errors
\be
P_{C}(\omega) =  {1/s \over (\omega/s)^2 + \pi^2}~,
    \label{nfanmkaacau}
\ee
with typical width $s$.
The Cauchy distribution (\ref{nfanmkaacau}) is one of the stable
L\'{e}vy distribution and possesses a power law tail with exponent
$\mu=1$ and a scale factor $C= s$.
Let us assume that the first (resp. second) sample $\xf$
(resp. $\xo$) has an error $\omega\spf$ (resp. $\omega\spo$)
distributed according to the Cauchy distribution (\ref{nfanmkaacau})
with typical width $s\spf$ (resp $s\spo$).
Then, the resulting distribution of the errors on $\xh$ is
of the same form (\ref{nfanmkaacau}) with the same exponent
$\mu=1$ while the scale factor is given by
\be
C_{\xh} = (1-\ko) s\spf + \ko s\spo~.  \label{fblabfva}
\ee
As we have found above in (\ref{fhafaa}), the weight $\ko$ that
minimizes $C_{\xh}$ is
\ba
\ko = \left\{ \begin{array}{lll}
0 & \qquad & \mbox{for $s\spo> s\spf$} \\
1 & \qquad & \mbox{for $s\spo< s\spf$} \end{array}
	\right. , \label{fhfagafaa}
\ea
since the Cauchy distribution is on the borderline $\mu=1$. This can be
verified
straightforwardly as a result of the linear dependence of $C_{\xh}$ on
$\ko$ in (\ref{fblabfva}) for which the optimization always selects one
of the boundaries.

Consider the case where $s\spf=1$ and $s\spo=2$.
The L\'{e}vy estimator imposes to choose $\ko = 0$,
i.e. to reject the information provided by the second sample $\xo$.
Let us now compare this recipe with the result obtained by applying
the standard Gaussian weight $K^{G}$ on data  generated by using
the two Cauchy laws with $s\spf=1$ and $s\spo=2$.
Specifically, we generated two sets of 1000 random numbers
$\omega\spf$ and $\omega\spo$, distributed according to the Cauchy law
(\ref{nfanmkaacau}) with $s\spf=1$ and $s\spo=2$.
  From each of these 1000 numbers, we can estimate numerically the
variance and find $(\sigma\spf)^2 = 3.36\times 10^5$ and
$(\sigma\spo)^2 = 4.07\times 10^5$.
The Gaussian estimator (\ref{qqgarwerqat}) then recommends the value
$K^{G}=[1+(\sigma\spo/\sigma\spf)^{2}]^{-1}\approx 0.45$ which is
very different from $\ko=0$ given by (\ref{fhfagafaa}).
We should stress that the estimations of the variances
$(\sigma\spf)^2$ and $(\sigma\spo)^2$ are highly unreliable because
they can change by orders of magnitude from one sample to another.
The reason is, as we have said, that the variance is mathematically
infinite in this case, and therefore any estimation of it is bound to
be dominated by the few largest random numbers that occur by chance in
the series.

Two lessons are thus to be learned from these numerical simulations:
i) estimating the variance for distributions with exponent $\mu <2$
leads to very unstable results;
ii) the resulting recommendation of the standard Gaussian
weight $K^{G}$ can be very wrong.

\section{Multivariate estimator}
\label{sec:multi}
\subsection{Definition of the model}
\label{sec:multi-model}

When the state variables are multi-dimensional, their errors may be
mutually dependent.
If the errors are distributed according to the power or L\'{e}vy laws,
we need to transform the coordinate of errors to express them as
linear sums of independent noises in order to be able
use the rules stated above in points (i)-(iv).

We consider a problem of estimating a multi-dimensional state vector
$\vx\in\bbR^{N}$ using two  samples $\vx\spf\in\bbR^{N}$ and
$\vx\spo\in\bbR^{N}$.
Both forecast and observations are made for all state variables.
As in the case for the univariate estimation, we assume that the estimate
$\vxh$ of $\vx$ can be expressed as a linear combination of $\vx\spf$
and $\vx\spo$.
Requiring the unbiased condition leads to one unknown weight
matrix in the estimation
\be \label{eq:vxh}
\vxh=(\mI-\mK\spo)\vx\spf+\mK\spo\vx\spo~,
\ee
where $\mI$ is an identity matrix and $\mK\spo\in\bbR^{N\times N}$ is the
weight matrix for the observation sample $\vx\spo$.
Our goal is to determine the optimal $\mK\spo$ which gives the
least uncertainty in $\vxh$.
We use the notation $\mK$ for the weight matrix in connection to the
standard Kalman(-Gaussian) gain matrix of sequential estimation \cite{ICGL}.

We assume that the sample error vectors are linear sums of $N$
independent $\mu$-variables with symmetric distributions
\be \label{eq:veta}
  \vep\spfo=\vx\spfo-\vx = \mG\spfo \vomg\spfo~
\ee
where the probability distribution of each $\omega\spfo_{l}$ is associated
with the same exponent $\mu$ and usually different individual scale factors
$C\spfo_{l}$. The assumption that the distributions of $\vomg\spfo$'s are
symmetric implies that the same scale factors $C\spfo_{l}$ characterize the
tail of large positive and negative realizations.
The transformation between independent $\vomg\spfo$ to mutually dependent
$\vep\spfo$ is provided by the matrix $\mG\spfo\in\bbR^{N\times N}$.
We shall use this decomposition scheme repeatedly in the sequel as it allows us
to treat in a simple way the interplay between
the power law distributions and the dependence between variables.

In the standard linear least-variance estimation theory, one 
calculates the covariance
matrix
${\hat \mP} \equiv\langle \vep \vep\spT \rangle$ of the error
$\epsilon = {\hat x} -x$ and minimizes the expectation of the distance between
$\vxh$ and $\vx$
in the mean-square sense, i.e., one minimizes the trace of the
covariance matrix ${\hat \mP}$. The covariance calculation is an essential
step to guarantee
that all error components are suitably accounted for.
We thus propose our key idea to generalize the covariance matrix in the regime
$\mu < 2$ where it does not exist by using the concept of the tail-covariance
defined as the matrix of scale factors of the distribution of all products
$\epsilon\spfo_{i}\epsilon\spfo_{j}$. We prefer this approach to the so-called
``co-variation'' \cite{SamorodinskyTaqqu}
as it is more intuitive and also presents nicer properties, in particular the
tail-covariance matrix remains symmetric. The intuitive meaning
of the co-variation is less
transparent than for the tail-covariance, which explicitly measures the
correlations between large events only, while the co-variation picks up
contributions from the core of the distributions.
Let us mention that a simplified version of the tail-covariance without signs
(see below) has been used in
the context of portfolio theory \cite{IJTAP}.

Consider the sample error vectors $ \vep\spfo$. We thus study the product
$\epsilon\spfo_{i}\epsilon\spfo_{j}$, whose probability distribution
constitutes the natural generalization of the covariance as already pointed
out.
Using properties (iii) and (iv) above,
one finds the following result, expressed
symbolically and without the superscripts for simplicity:
\be \label{eq:epep}
  \epsilon_{i}\epsilon_{j}
   \simeq \sum_{l}^{N} |G_{il}| |G_{jl}| \ [\mbox{${\mu\over 2}$-variable}]
   + \sum_{l'}^{N}\sum_{l''\neq l'}^{N} |G_{il'}| |G_{jl''}|
	\ [\mbox{${\mu}$-variable}]~,
\ee
where the symbol ``$\mu$-variable'' defines a random variable distributed
with a density distribution with a power law tail of exponent $\mu$.
Expression (\ref{eq:epep})
  means that the tail of the distribution of the product
$\epsilon_{i}\epsilon_{j}$ is {\it dominated by the first term} which has
the smallest exponent $\mu/2$ and thus heaviest tail, and
will directly be sensitive to the product $|G_{il}| |G_{jl}|$ for
all the independent errors $\omega_{l}$.
More precisely, an analysis of the cumulative distribution
of the products $\epsilon_{i}\epsilon_{j}$
will give an asymptotic slope of $-{\mu \over 2}$ in a log-log plot 
and a scale factor
$C_{l}$
associated with $\omega_{l}$  proportional to
$|G_{il}|^{\mu} |G_{jl}|^{\mu}$.

We use this set of scale factors associated with the distributions of
$\epsilon_{i}\epsilon_{j}$ in order to
define the tail-covariance matrix $\mB\in\bbR^{N\times N}$ with the
following guidelines.
It is natural that the tail-covariance matrix should contain the
information on the tail of
the products $\epsilon_{i}\epsilon_{j}$ with distribution (\ref{eq:epep}).
A bona-fide generalization of the covariance matrix requires two additional
conditions.
It should be sensitive to the sign of the dependence between the variables,
i.e.,
if $\epsilon_{i}$
increases (resp. decreases) on average conditioned on the increase of
$\epsilon_{j}$,
the dependence is positive (resp. negative), generalizing
the existence of positive and negative correlations. In addition, a
suitable definition of the
tail-covariance matrix should be such that it recovers the standard
covariance matrix
for the value $\mu=2$ corresponding to the special case where the stable laws
reduce to the Gaussian distribution, as briefly recalled in Appendix 
\ref{Appendix:levy}.
These considerations lead to the following unique specification for the
tail-covariance
matrix by diagonalization:
\be \label{eq:B}
  \mB=\mG\pL{{\mu\over 2}} \mC \mG\pLT{{\mu\over 2}}~.
\ee
where $\mC\in\bbR^{N\times N}$
is a diagonal matrix associated with $\vomg$
and $\mG\pL{{\mu\over 2}}\in\bbR^{N\times N}$ is the corresponding 
eigenvector matrix.
The operator $\{\cdot\}\pL{\beta}$ means that each element of matrix or
vector is defined by
\be
\mG\pL{{\beta}}_{ij} = {\rm sign}(G_{ij})~|G_{ij}|^{\beta}~,  \label{bjssz}
\ee
i.e., the absolute value of each element is raised to the power $\beta$ and
then multiplied by its sign.
This operation can be applied to scalars as well.
Without the sign operator in (\ref{bjssz}), the tail-covariance matrix $\mB$
would be just the matrix of scale factors of all the products
$\epsilon_{i}\epsilon_{j}$.
The introduction of the sign function is an essential additional ingredient
introduced to account for the sign of the dependence between the variables.

For $\mu=2$, $\mG\pL{{\mu \over 2}}_{ij}
={\rm sign}(G_{ij})~|G_{ij}|= G_{ij}$, and we check that
the tail covariance is exactly the same as the
error covariance
\be \label{eq:BP}
  \mB=\mtrx{P}\equiv\langle \vep \vep\spT \rangle~.
\ee
This correspondence may appear paradoxical if one interprets the case $\mu=2$
as corresponding to a power law, which has infinite
variance. As we already discussed for the unidimensioncal case,
the technical reason for this comes
>from the form (\ref{bbvvvcnc})
of the characteristic function $\hat{L}_{\mu}(k)$ of a distribution with
a power law tail, as given in \cite{Feller}
(Problem 15 of Section 12) or of a symmetric L\'evy laws.
We stress again
the important point that, for $\mu$ strickly less than $2$,
the inverse Fourier transform of $\hat{L}_{\mu}(k)$ defined by
(\ref{bbvvvcnc}) gives a power law tail,
while for $\mu=2$, it gives a Gaussian law. The continuity between
the expressions (\ref{eq:B}) and (\ref{eq:BP}) can thus be traced back
to that of $\hat{L}_{\mu}(k)$ as a function of $\mu$ at $\mu=2$.

The transformation of the scale factors between the mutually dependent
errors $\vep$ and independent errors $\vomg$  as in (\ref{eq:B})
can be performed in both directions,  i.e., not only from right-
to left-hand side to compute $\mB$ when $\mG$ and $\mC$ are known,
but also from left- to right-hand side to obtain $\mC$ (and $\mG$)
by diagonalization of $\mB$.

\subsection{Solution}
\label{sec:multi-solution}

Our aim is to obtain the optimal weight $\mK\spo$ in (\ref{eq:vxh})
for the  best estimate  $\vxh$. In this goal, we form the set of products
$\epsilon_{i}\epsilon_{j}$ where $\epsilon = {\hat x} -x$ and study their
probability distribution. As in (\ref{eq:epep}) and (\ref{eq:B}), we retain
only
the term decaying as a power law with an exponent $\mu/2$ and get the following
tail covariance matrix of ${\hat \vx}-\vx$  for an
arbitrary weight matrix $\mK\spo$:
\be \label{eq:Bh}
{\hat \mB}=
  \left(\mG\spf-\mK\spo\mG\spf\right)\pL{{\mu\over 2}}\mC\spf
  \left(\mG\spf-\mK\spo\mG\spf\right)\pLT{{\mu\over 2}} +
  \left(\mK\spo\mG\spo\right)\pL{{\mu\over 2}}\mC\spo
  \left(\mK\spo\mG\spo\right)\pLT{{\mu\over 2}}~.
\ee
Here, we assume that errors $\vep\spf$ and $\vep\spo$ are mutually
independent.

In the univariate case, the optimization process is unique and corresponds to
minimizing ${\hat C}$ with respect to $\ko$ as in (\ref{jfjks}).
In the multivariate case, however, the optimization may be defined
in several ways.
For example, as recalled above,
the standard linear least-variance estimation theory
attempts
to minimize the expectation distance between $\vxh$ and $\vx$
in the mean-square sense, i.e., to minimize $\trace \  {\hat \mP}$
given by (\ref{eq:BP}).
For $\mu<2$ where the
covariance ${\hat \mP}$ does not exist, we propose to minimize
the ``average'' scale factor, i.e., $\trace \  {\hat \mB}$.
Such an optimization implies that the uncertainty in
$\vxh$ is globally the smallest (see Section
2-2 for the univariate case).

Since the expression (\ref{eq:Bh}) is not smooth in $\ko$ due to the presence
of the absolute values, some care must be taken in the minimization.
The non-smooth character of (\ref{eq:Bh})
makes the differentiation approach to the minimization more cumbersome, as one
must keep track of the discontinuities.
The optimal $\mK\spo$ is obtained by solving
\be \label{eq:BhA}
  \pdrv{}{\mK\spo} \trace \  {\hat \mB}  = 0~,
\ee
where
\ba  \label{eq:Bhii}
  {\hat B}_{ii}&=&\sum_{p=1}^{N}
  \left( G\spf_{ip}-\sum_{m=1}^{N}K\spo_{im}G\spf_{mp}\right)
   \pL{{\mu \over 2}}  C\spf_{p}
  \left( G\spf_{ip}-\sum_{m=1}^{N}K\spo_{im}G\spf_{mp}\right)
   \pL{{\mu \over 2}}  \nonumber \\
  && + \sum_{p=1}^{N}
  \left(\sum_{m=1}^{N}K\spo_{im}G\spo_{mp}\right)\pL{{\mu \over 2}}
   C\spo_{p}
  \left(\sum_{m=1}^{N}K\spo_{im}G\spo_{mp}\right)\pL{{\mu \over 2}}
  \nonumber \\
  &=&\sum_{p=1}^{N}
  \left| G\spf_{ip}-\sum_{m=1}^{N}K\spo_{im}G\spf_{mp}\right|^{\mu}
   C\spf_{p}  + \sum_{p=1}^{N}
  \left|\sum_{m=1}^{N}K\spo_{im}G\spo_{mp}\right|^{\mu} C\spo_{p}~.
\ea
Taking the derivative with respect to $K\spo_{ij}$,
we obtain:
\ba
\pdrv{}{K\spo_{ij}} \trace \  {\hat \mB}
  &=&\pdrv{}{K\spo_{ij}} {\hat B}_{ii}\nonumber \\
  &=& \mu \sum_{p=1}^{N}  \left[ -
  \left(G\spf_{ip}- \sum_{m=1}^{N}K\spo_{im}G\spf_{mp}\right)\pL{\mu-1}
   G\spf_{jp}C\spf_{p} \nonumber \right. \\
  & & \qquad +  \left.
   \left(\sum_{m=1}^{N}K\spo_{im}G\spo_{mp}\right)\pL{\mu-1}
   G\spo_{jp}C\spo_{p}\right] =0~,\label{eq:dtrBdAij}
\ea
where we have used the fact that $d|x|^{\mu}/dx=\mu {\rm sign}(x)~|x|^{\mu-1}
= \mu x^{[\mu-1]}$, with our definition (\ref{bjssz}).
Note that the same $N$ coefficients $K\spo_{im}$ with $m=1,...,N$ and only them
occur in the $N$ equations $\pdrv{}{K\spo_{ij}} \trace \  {\hat \mB}$
obtained by fixing $i$ and varying $j$ from $1$ to $N$.
Finding the optimal $\mK\spo$ thus involves solving $N$ independent systems
of equations, one for each individual diagonal term ${\hat B}_{ii}$, where
each system consists of the $N$
nonlinear equations $\pdrv{}{K\spo_{ij}} {\hat B}_{ii}=0$ for
the $N$ unknowns  $K\spo_{ij}$ at $i$ fixed.

Optimality of $\mK\spo$ is ensured by the positive
definiteness of the Hessian matrix for ${\hat B}_{ii}$
with respect to $K\spo_{ij}$, for each of the $N$ independent systems (each
defined as we have shown by a fixed $i$) of $N$ equations (obtained 
by varying $j$ at
fixed $i$). While for the univariate case, this amounts to ensure the 
validity of only
one equation
(\ref{eq:optko}), in the multivariate case we need to study the 
positive definiteness
of the Hessian:
\ba\label{eq:ddtrBddAij}
  \pdrv{^{2}}{(K\spo_{ij})\partial(K\spo_{iq})}
  {\hat B}_{ii} = &
  \mu (\mu-1) & \left[\sum_{p=1}^{N} (G\spf_{jp})(G\spf_{qp})
  \left|G\spf_{ip}- \sum_{m=1}^{N}K\spo_{im}G\spf_{mp}\right|^{\mu-2}
   C\spf_{p}  \right. \nonumber \\
  & & \left. + \sum_{p=1}^{N }(G\spo_{jp})(G\spo_{qp})
   \left|\sum_{m=1}^{N}K\spo_{im}G\spo_{mp}\right|^{\mu-2}
   C\spo_{p}\right]~,
\ea
expressed at the values of $\mK\spo$ solving (\ref{eq:dtrBdAij}).
If this matrix is positive definite, then $K\spo_{ij}$ minimizes
${\hat B}_{ii}$ and is therefore optimal.
There are two issues associated with this formulation for the
optimal $\vctrg{\kappa}$ (i.e., $\mK\spo$): 1) the possible existence of
singular terms for $\mu<2$;
and 2) the solvability of the nonlinear system. We address them in the
appendix \ref{issues} and show that the matrix is indeed positive definite
and that there is only one solution.

\subsection{Special cases}
\label{sec:multi-special}

\subsubsection{Case $\mu=2$}

For $\mu=2$, the L\'{e}vy estimator is the same as the Gaussian one
which minimizes $\trace \  {\hat \mP}=\trace \  {\hat \mB}$ as given by
(\ref{eq:BP}).
In this case, the system becomes perfectly linear with
${\mu\over 2}=\mu-1=1$ and leads to the following analytical solution for
the optimal weight given in matrix form
\be
  \mK^{G}=\mB\spf \left(\mB\spf+\mB\spo\right)^{-1}~, \label{eq:Ah}
\ee
which results in the optimal estimates of the state variable and
corresponding covariance matrix
\ba \label{eq:vxhKG}
  {\hat \vx}^{G}&=& \mB\spo \left(\mB\spf+\mB\spo\right)^{-1}\vx\spo
       +  \mB\spf \left(\mB\spf+\mB\spo\right)^{-1}\vx\spf~,\\
  {\hat \mB}^{G}&=&
   \mB\spo \left(\mB\spf+\mB\spo\right)^{-1}\mB\spf~. \label{eq:mBhKG}
\ea

\subsubsection{Independent noise}

When the errors $\vep\spfo$ are independent, i.e., $\mG\spfo=\mI$,
the trace and diagonal components (\ref{eq:Bhii}) for
the average scale factor can be simplified into
\be  \label{eq:Bhii-uni}
  \trace \ {\hat \mB}
   =\sum_{i=1}^{N} \sum_{p=1}^{N}\left[ \left(1-K\spo_{ip}\right)^{\mu}
C\spf_{p}
   +\left(K\spo_{ip}\right)^{\mu} C\spo_{p}\right]~,
\ee
where we have omitted the absolute values and the
sign functions as we look for values of the Kalman
weights between $0$ and $1$.
Therefore, the problem can be reduced to the univariate estimate for each
element $x_{i}$ independently, as defined in equation (\ref{faohgal}).
The equations (\ref{jfjks},\ref{eq:optko}) are thus replaced by their exact
equivalent derived from the two conditions (\ref{eq:BhA}) and
(\ref{eq:ddtrBddAij}).
The solution  $K\spo_{ii}$ for each $i$
is given by (\ref{qqwersfqt}) where $K\spo_{ii}=\ko$ is obtained by
setting $\lambda=(C\spo_{i})^{1\over\mu}/(C\spf_{i})^{1\over\mu}$
(see (\ref{fbbrfjkfwq})). Due to the hypothesis of noise independence,
the non-diagonal coefficients $K\spo_{ij}$ are all zero for $i\neq j$.

\section{The Kalman-L\'{e}vy filter}
\label{sec:KLF}

\subsection{Problem}
\label{sec:KLF-problem}

We are now in a position to construct a sequential data assimilation
methodology when the noises are distributed according to the power or
L\'{e}vy law. The standard assimilation problem is formulated as follows.
We consider a linear discrete stochastic dynamical system
of state variables $\vx\in\bbR^{N}$
\be
\vxk\spt =\mMkkn \vxkn\spt + \vet\sbkn\spt~, \label{eq:xt}
\ee
where superscript $\{\cdot\}\spt$ denotes the true state
and $\vet\sbkn\spt$ is the dynamical noise.
The index $k$ corresponds to the time sequence when
the observations $\vyk\spo\in\bbR^{L\sbk}$ are taken as
\be
\vyk\spo= \mHk \vxk\spt  + \vep\sbk\spo~,
\ee
where $\vep\sbk\spo$ is the observational noise and
$\mHk\in\bbR^{L\sbk\times N}$ is the linear observation function
which can vary at each time step $k$. These observations $\vyk\spo$
are assumed to be linear functions of the state variable $\vxk\spt$ of the
system with an additive noise.

The estimation methodology developed in the previous sections is now
extended to perform a filtering in order to estimate $\vx\spt$ sequentially,
by assimilating $\vyk\spo$ into the deterministic forecast $\vxk\spf$.
Here, we assume that $\mMkkn$ and $\mHk$ are known.
The assimilation cycle $k$ is defined over a time interval
$[k-1,k]$ between the two adjacent observation events. It
consists of the following two steps:
\begin{enumerate}
\item a deterministic forecast $\vxk\spf$ of $\vxk\spt$ from a
given initial condition $\vxkn\spa$ as the best estimate of
$\vxkn\spt$ based on the model
\be
\vxk\spf=\mMkkn \vxkn\spa~. \label{eq:Mxa}
\ee
The forecast is based on the analysis performed at the previous time step.

\item This forecast is then used to construct the new analysis
$\vxk\spa$ which is mixed with the assimilated observation. This leads to
the probabilistic analysis $\vxk\spa$ of $\vxk\spt$ obtained as the weighted
average of $\vxk\spf$ and $\vyk\spo$ under the unbias assumption
at time $k$:
\be
\vxk\spa= \vxk\spf + \mKk \left( \vyk\spo - \mHk \vxk\spf \right) =
\left(\mI-\mKk\mHk\right) \vxk\spf + \mKk \vyk\spo~.
\ee
\end{enumerate}
Accordingly, the errors associated with $\vxk\spf$ and $\vxk\spa$
are auto-regressive processes.
\ba
\vxk\spf - \vxk\spt &=& \mMkkn \left( \vxkn\spa-\vxkn\spt \right)
- \vet\sbkn\spt~,    \label{eq:error-forecast} \\
\vxk\spa - \vxk\spt &=& \left(\mI-\mKk\mHk\right)
  \left(\vxk\spf - \vxk\spt \right) + \mKk \vep\sbk\spo \nonumber \\
  &=& \left(\mI-\mKk\mHk\right) \mMkkn \left(\vxkn\spa-\vxkn\spt\right)
  - \left(\mI-\mKk\mHk\right)\vet\sbkn\spt + \mKk \vep\sbk\spo
   \label{fhghghnfg}
  \label{eq:error-analysis}
\ea
and the only unknown to be determined is the so-called ``gain matrix''
$\mKk$ needed in order to complete the assimilation cycle.

Our goal is therefore to determine the gain matrix $\mKk^{L}$, where the
superscript $L$ refers to the Kalman-L\'evy filter, which
results in the least uncertainty in $\vxk\spa$ in each assimilation cycle
when the noises are distributed according to the power or L\'{e}vy law
with the exponent $\mu$. As in (\ref{eq:veta}),
we express the sample error vectors as linear sums of $N$
independent $\mu$-variables
\be
\vet\sbk \equiv \mGk\spet \vomgk\spet, \qquad
\vep\sbk \equiv \mGk\spep \vomgk\spep~. \label{eq:etepGw}
\ee
Because (\ref{eq:error-forecast})
and (\ref{fhghghnfg}) define linear autoregressive processes
with L\'evy-stable or power law probability distribution of
the noises,  the errors in forecast and
analysis are also distributed according to the power or L\'{e}vy
laws with the same exponent $\mu$. Without loss of generality, they can
thus be written as
\be
\vxk\spfa - \vxh\sbk \equiv\mGk\spfa \vomgk\spfa~, \label{eq:xGw} \\
\ee
which defines the matrices $\mGk\spfa$ and the vectors $\vomgk\spfa$ of
independent L\'evy or power law processes.
Consequently, all error and noise distributions in the
assimilation cycles are characterized by the corresponding tail
covariance matrices
\ba
\mBk\spfa&\equiv&
   \left(\mGk\spfa\right)\pL{{\mu\over 2}} \mCk\spfa
   \left(\mGk\spfa\right)\pLT{{\mu\over 2}} \label{eq:Bfa} \\
\mBk^{\eta,\epsilon}&\equiv&
   \left(\mGk^{\eta,\epsilon}\right) \pL{{\mu\over 2}}\mCk^{\eta,\epsilon}
   \left(\mGk^{\eta,\epsilon}\right)\pLT{{\mu\over 2}}~, \label{eq:Betep}
\ea
where $\mC\in\bbR^{N\times N}$
is a diagonal scale-factor matrix associated with $\vomg$
and $\mG\pL{{\mu\over 2}}$ is the corresponding eigenvector matrix.
Given a $\mB$, the corresponding $\mG$ and $\mC$ can be obtained by
diagonalization.

\subsection{Solution}
\label{sec:KLF-solution}

The optimal Kalman-L\'{e}vy (KL) filter $\mKk^{L}$ which  minimizes the
global average error
is obtained by minimizing the trace of the tail-covariance  matrix
$\mBk\spa$ of the resulting probabilistic analysis $\vxk\spa$ of $\vxk\spt$.
Since the expression for $\mBk\spa$ is not smooth in $\mKk^{L}$ due to
the presence of the absolute values, some care must be taken in the
minimization as already discussed. The non-smooth character of $\mBk\spa$
makes the
differentiation approach to the minimization more cumbersome, as
one must keep track of the discontinuities. However, we use the approach
described in Appendix B to check if the
two conditions are simultaneously verified:
\be \label{eq:A}
  \pdrv{}{\mKk^{L}} {\rm trace} \mBk\spa = 0~,
\ee
and positive-definiteness of the Hessian matrix
\be
\pdrv{^2}{(\mKk^{L})^{2}} {\rm trace} \mBk\spa ~.
\ee
We solve the first condition and then examine the second condition.
This is achieved by taking the following two steps
in each assimilation cycle.

\newcounter{stp}
\begin{list}
{Step \arabic{stp}.}{\usecounter{stp}
  \setlength{\rightmargin}{\leftmargin}}
\item Dynamic forecast: \\
Given a set of initial conditions described by the subscript
$\{\cdot\}\sbkn$ which are known, the forecast is performed deterministically
to advance from $k-1$ to $k$ based on (\ref{eq:Mxa}) and
(\ref{eq:error-forecast})
\be
\vxk\spf = \mMkkn \vxkn\spa \label{eq:xkf}
\ee
leading to the tail-covariance of the forecasts at time $k$:
\be
\mBk\spf = \left(\mMkkn\mGkn\spa\right)\pL{{\mu\over 2}} \mCkn\spa
\left(\mMkkn\mGkn\spa\right)\pLT{{\mu\over 2}} + \mBkn\spet~, \label{eq:Bkf}
\ee
where the definition (\ref{eq:B},\ref{bjssz}) is used.

\item Probabilistic analysis: \\
Given the forecast $\vxk\spf$ with $\mBk\spf$ from Step 1 along with
the observations $\vyk\spo$ with tail-covariance $\mBk\spep$,
the analysis provides the optimal estimate
\be
\vxk\spa = \vxk\spf+\mKk^{L}\left(\vyk\spo-\mHk\vxk\spf\right)~
        \label{eq:xka}
\ee
with tail-covariance
\ba
\mBk\spa&=&\left(\mGk\spf- \mKk^{L}\mHk\mGk\spf\right)\pL{{\mu\over 2}}
  \mCk\spf \left(\mGk\spf- \mKk^{L}\mHk\mGk\spf\right)\pLT{{\mu\over 2}}
  \nonumber \\
  & &  + \left(\mKk^{L}\mGk\spep\right)\pL{{\mu\over 2}} \mCk\spep
    \left(\mKk^{L}\mGk\spep\right)\pLT{{\mu\over 2}}~, \label{eq:Bka}
\ea
where the definition (\ref{eq:B},\ref{bjssz}) is used.
By letting subscripts represent the matrix elements
and dropping the time index $k$ for simplicity,
we solve for the optimal filter $\mK^{L}$ so that it satisfies
the conditions (\ref{eq:A}).
The diagonal elements of $B\spa$ can be explicated as follows:
\ba  \label{eq:Bii}
  B\spa_{ii} &=&\sum_{p=1}^{N}
  \left|G\spf_{ip}- \sum_{m=1}^{L}K_{im}H\spG_{mp}\right|^{\mu}
	 C\spf_{p} + \sum_{q=1}^{L}
  \left|\sum_{m=1}^{L}K_{im}G\spep_{mq}\right|^{\mu} C\spep_{q}~,
\ea
with
\be
\mH\spG\equiv\mH\mG\spf~.
\ee

Because $B\spa_{ii}$ does not depend on $K_{qj}$ for $q\neq i$,
the first condition for the optimal $\mKk^{L}$  (\ref{eq:A}) is
\ba \label{eq:KLopt}
\pdrv{}{K^{L}_{ij}} \trace \  \mB\spa
  &=&\pdrv{}{K^{L}_{ij}}  B\spa_{ii} =0 \nonumber \\
  &=& \mu \left[ - \sum_{p=1}^{N}
  \left(G\spf_{ip}- \sum_{m=1}^{L}K^{L}_{im}H\spG_{mp}\right)\pL{\mu-1}
   H\spG_{jp}C\spf_{p} \right. \nonumber \\
  & & +  \left. \sum_{q=1}^{L}
  \left(\sum_{m=1}^{L}K^{L}_{im}G\spep_{mq}\right)\pL{\mu-1}
   G\spep_{jq}C\spep_{q}\right] =0 ~,
\ea
using the signed power operator as in the case of
(\ref{eq:dtrBdAij}).

For the optimal filter $\mK^{L}$ so obtained, positive definitiveness of
the Hessian matrix
\ba \label{eq:KLmin}
  \pdrv{^{2}}{(K^{\rm L}_{ij})\partial(K^{\rm L}_{iq})}  \trace \  \mB\spa
  &=&  \pdrv{^{2}}{(K^{\rm L}_{ij})\partial(K^{\rm L}_{iq})} 
B\spa_{ii} \nonumber \\
  &=& \mu (\mu-1)  \left[\sum_{p=1}^{N}
  \left|G\spf_{ip}- \sum_{m=1}^{L}K^{L}_{im}G\spep_{mp}\right|^{\mu-2}
   (H\spG_{jp}) (H\spG_{qp}) C\spf_{p} \right. \nonumber \\
  & & + \left. \sum_{q=1}^{L}
  \left|\sum_{m=1}^{L}K^{L}_{im}G\spep_{mq}\right|^{\mu-2}
   (G\spep_{jp}) G\spep_{qp}C\spep_{q}\right]~,
\ea
is satisfied for $\mu>1$, following the same approach as discussed
in Section 3-2 and in Appendix B.

The optimal estimate $\vxk\spa$ and tail-covariance $\mBk\spa$ obtained by
substituting
$\mKk^{L}$ into (\ref{eq:xka}) and (\ref{eq:Bka}) become a
set of initial conditions for the next assimilation cycle $k+1$.
\end{list}

Similarly to the case of the multivariate estimator which is solution of
(\ref{eq:dtrBdAij})
discussed in Section \ref{sec:multi-solution},
the first condition for the optimal KL filter (\ref{eq:KLopt}) for a fixed
$i$ leads  to a set of $N$  self-contained nonlinear equations for
$N$ unknowns for $j=1,\ldots,N$.
Minimization of the average scale factor is therefore equivalent to the
minimization of each tail-covariance element $B\spa_{ii}$ for
$x\spa_{i}-x\spt_{i}$ with respect to the elements of $\mKk$ with at least
one of the
indexes equal to $i$.
Such a set of solutions of the nonlinear equations for an arbitrary $\mu$
may not be available analytically but can be obtained numerically.
For $\mu=2$, the KL filter is reduced to the same formula as the
conventional Kalman-Gaussian (KG) filter,
i.e., the forecast error tail-covariance (\ref{eq:Bkf}) is
\be
\mBk\spf=\mMkkn\mBkn\spa\mMkkn\spT + \mBkn\spet~.
\ee
The analysis error tail-covariance  (\ref{eq:Bka}) is
\ba
\mBk\spa&=&
  \left(\mI- \mKk^{G}\mHk\right) \mBk\spf~,
\ea
where the optimal KG gain that satisfies (\ref{eq:A}) is given by
\be \label{eq:AKG}
\mKk^{G}=\mBk\spf\mHk\spT
\left(\mHk\mBk\spf\mHk\spT+\mBk\spep\right)^{-1}~.
\ee
The analysis state variable
$\vxk\spa$ is also obtained by substituting (\ref{eq:AKG})
into (\ref{eq:xka}).
In this case, sequential data assimilation does not require
any diagonalization of the tail covariance matrix.

\subsection{Univariate Kalman-L\'{e}vy filter}
\label{sec:KLF-uni}

\subsubsection{Solution}
\label{sec:KLF-uni-solution}

To understand the fundamental  properties of the
KL filter, we study the univariate problem
with $N=1$ and $L=1$ in detail for the case where the exponent $\mu$ is
larger than $1$. The case $\mu \leq 1$ has been discussed in section 2 and
leads
to a Kalman weight equal to either $0$ or $1$.

In this case, the tail-covariances $B^{{\rm f,a,t}}$ correspond to the scale
factors $C^{{\rm f,a,t}}$
directly and we have the following data assimilation cycle.
\newcounter{stpuni}
\begin{list}
{Step \arabic{stpuni}.}{\usecounter{stpuni}
  \setlength{\rightmargin}{\leftmargin}}
\renewcommand{\theenumi}{Step \arabic{enumi}}
\item Dynamic forecast:
\be
\xf\sbk = \Mkkn\xa\sbkn \label{eq:uni-xkf}
\ee
with tail-covariance matrix
\be
\Bk\spf = |\Mkkn|^{\mu} B\sbkn\spa + B\sbkn\spet \label{eq:uni-Bkf}
\ee
as derived from (\ref{eq:error-forecast}) using the calculation rules given
above in points (i)-(iv).

\item Probabilistic analysis: \\
\be
\xa\sbk = \left(1 - K_k H_k\right) \xf\sbk + K_k y\sbk\spo~,
\ee
leading to the following scale factor
\be \label{eq:uni-BkaK}
\Bk\spa =  \left|1 - K_k H_k\right|^{\mu} \Bk\spf +
|K_k|^{\mu} \Bk\spep~.
\ee
Its minimization with respect to $K_k$ leads to
\ba
\xa\sbk&=&{(\lambda\sbk^{H})^{{\mu\over \mu-1}} \over
    1 + (\lambda\sbk^{H})^{{\mu\over \mu-1}}} \xf\sbk
  + {{H\sbk}^{-1} \over 1+(\lambda\sbk^{H})^{{\mu\over \mu-1}}} y\sbk\spo
    \label{eq:uni-xa}\\
\Bk\spa&=&{(\lambda\sbk^{H})^{\mu} \over
  \left[1+(\lambda\sbk^{H})^{{\mu\over \mu-1}}\right]^{\mu-1}}
    \Bk\spf~, \label{eq:uni-Bka}
\ea
after substituting the optimal KL gain
\be
  K\sbk^{L} = {{H\sbk}^{-1} \over 1+(\lambda\sbk^{H})^{{\mu\over \mu-1}}}~,
  \label{eq:uni-KkL}
  \ee
  with the modified relative error ratio
  \be
  \lambda\sbk^{H}\equiv{\lambda\sbk\over H\sbk} =
  {(\Bk\spep)^{{1\over\mu}} \over H\sbk (\Bk\spf)^{{1\over\mu}}}~.
   \label{eq:uni-lambda}
\ee
Notice that expression (\ref{eq:uni-Bka}) is nothing but rewriting
(\ref{eq:CaL}).
\end{list}

\subsubsection{Properties of the solution}
\label{sec:KLF-uni-prop}

Because the KL filter is designed to minimize $\Bk\spa$,
we gain insight into its performance
by investigating $\Bk\spa$ along with $\Bk\spf$
in each assimilation cycle.
While the state variable $x\sbk\spfa$ has a stochastic dynamics
through the observation $y\spo$, their scale factors  $\Bk\spfa$
are completely deterministic.
Using  (\ref{eq:uni-Bka}) and (\ref{eq:uni-Bkf}), the evolution of
$\Bk\spfa$ can be expressed as uncoupled one-dimensional maps:
\ba \label{eq:uni-Bf}
\Bk\spf&=&{ |\Mkkn \lambda\sbkn^{H}|^{\mu} \over
\left[1+(\lambda\sbkn^{H})^{{\mu\over \mu-1}}\right]^{\mu-1}}
   B\sbkn\spf + B\sbkn\spet~ \\
\Bk\spa&=& { (\lambda\sbk^{H})^{\mu}  \over
\left[1+(\lambda\sbk^{H})^{{\mu\over \mu-1}}\right]^{\mu-1}}
  \left[ |\Mkkn|^{\mu}B\sbkn\spa + B\sbkn\spet \right]~ ,
\label{eq:uni-Ba}
\ea
where $\lambda\sbk^{H}$ in $\Bk\spa$ can be given in terms of
$B\sbkn\spa$ for (\ref{eq:uni-Ba}) by substitution of
(\ref{eq:uni-Bkf}) into (\ref{eq:uni-lambda}).

Two limiting cases can be analyzed.
  First, in the limit where the main origin of variability in the factor
$F$ multiplying $B\sbkn\spf$ in the r.h.s. of (\ref{eq:uni-Bf}) comes from
either
$\Mkkn$, $\Bk\spep$ or $H\sbk$ and not from $\Bk\spf$, this factor $F$
can be considered to be approximately independent of $B\sbkn\spf$.
The expression (\ref{eq:uni-Bf}) becomes a multiplicative noisy
auto-regressive equation which has been much studied in the literature
\cite{Kesten,Calan,Sorcont,Sormul}.
The most important result is that $\Bk\spf$ remains finite at all times
if the expectation of the logarithm of the factor $F$ multiplying
$B\sbkn\spf$ in (\ref{eq:uni-Bf}) is negative.
This condition ensures that $\Bk\spf$ does not grow exponentially at large
times.
Usually, in this regime, if the factor $F$ exhibits intermittent
excursions to values larger than one, it can be shown that the scale
  factors $B\sbkn\spf$ themselves will be distributed according to a
power law distribution.
A similar result holds for $\Bk\spa$, whose upper limit is bounded by
$\Bk\spf$.

The second interesting case occurs when the system is stationary,
i.e.,  $M=\Mkkn$, $H=H\sbk$, and
$B^{\eta,\epsilon}=\Bk^{\eta,\epsilon}$ for all $k$.
By defining nondimensional tail-covariances
(which are replaced by scale factors in this single variable case)
normalized by the dynamical error's scale factor $B\spet$,
\be \label{eq:uni-bkfa}
\bk\spfa={\Bk\spfa\over B\spet}~,
\ee
the corresponding dynamical maps (\ref{eq:uni-Bf}) and (\ref{eq:uni-Ba})
are reduced into
\ba
\bk\spf&=&{|M \lmdb|^{\mu} \over
  \left\{ 1 + \left[
   {\lmdb\over (\bkn\spf)^{{1\over \mu}}}\right]^{{\mu\over \mu-1}}
   \right\}^{\mu-1}}~ + 1~,  \label{jfjnnkkk} \\
\bk\spa&=&{(\lmdb)^{\mu} \over
  \left\{ 1 +  \left[
   {\lmdb\over (|M|^{\mu}\bkn\spa+1)^{{1\over \mu}}}\right]
    ^{{\mu\over \mu-1}} \right\}^{\mu-1}}~.
    \label{jajnnkkk}
\ea
For a given  $\mu$, the two parameters controlling the evolution of
$\bk\spfa$ are the dynamical coefficient $M$ and the
ratio of the characteristic error sizes of the dynamics over the
observation defined by
\ba \label{eq:uni-bH}
\lmdb &\equiv& {1\over H}
  {(B\spep)^{1\over \mu} \over (B\spet)^{1\over \mu}}~.
\ea
The corresponding KL gain is
\be \label{eq:uni-KL}
  K\sbk^{L} = {H^{-1} \over
   1 +  \left[
  {\lmdb\over (\bkn\spf)^{{1\over \mu}}}\right]^{{\mu\over\mu-1}}}~,
\ee
which can be expressed in terms of $b\sbkn\spa$ as well.
The KL filter parameter (\ref{eq:uni-bH}) and gain (\ref{eq:uni-KL})
are the counterpart of (\ref{fbbrfjkfwq}) and (\ref{qqwersfqt}),
and the resulting KL analysis (\ref{jajnnkkk}) takes the form similar
to (\ref{eq:CaL})
of the L\'{e}vy estimator in Section \ref{sec:uni-solution}.

For any values of $b\sbkn\spfa$, the scale factors $\bk\spfa$ at the next
time step are bounded:
\ba
1<&\bk\spf&< |M \lmdb|^{\mu} +1 \label{eq:bkfb} \\
{(\lmdb)^{\mu} \over \left[1+(\lmdb)^{{\mu\over \mu-1}}\right]^{\mu-1}}
< &\bk\spa& <(\lmdb)^{\mu}~, \label{eq:bkab}
\ea
indicating that the univariate sequential estimation system cannot
diverge as long as the exponent $\mu$ is finite.

\begin{figure}
\begin{center}
\epsfig{file=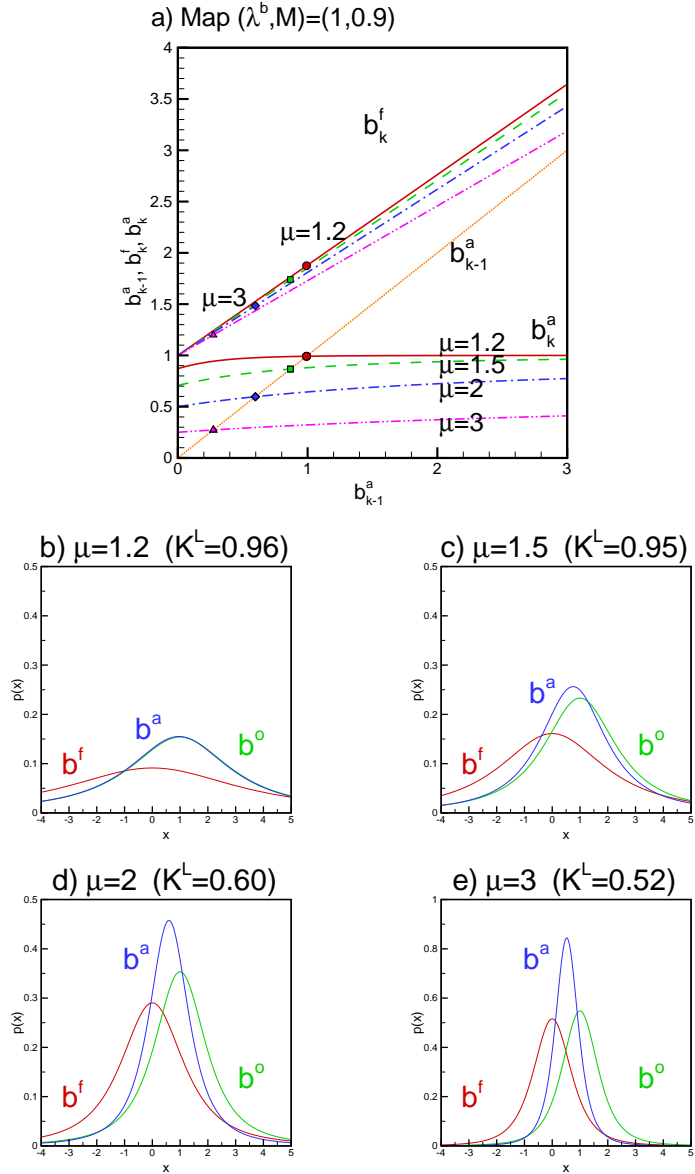,height=16cm}
\caption{\protect\label{fig:unimap}
a) Graphs of $\bk\spf$ and $\bk\spa$ as
function of $b\sbkn\spa$ for $(\lmdb,M)=(1,0.9)$.
For both $\bk\spf$ and $\bk\spa$,
the lines represent the corresponding maps and the symbols are the stable
fixed-points;
solid, dash, dash-dot, and dash-dot-dot lines, as well as
circle, square, diamond, and triangle symbols correspond
to $\mu=1.2$, 1.5, 2 and 3, respectively;
b) error probability distribution  for the stable fixed-point
corresponding to $\mu=1.2$ and parameters $(x\spf,y\spo)=(0,1)$ and
$H=1$, so that $x\spa=K^{L}$;
c, d and e) same as b but corresponding to $\mu=1.5$, 2, and 3,
respectively.}
\end{center}
\end{figure}

The evolution of the scale factors is demonstrated in Figure
\ref{fig:unimap}a.
The maps are represented by the graphs of $\bk\spf$ and $\bk\spa$ as
functions of $b\sbkn\spa$ for four values of
the exponent $\mu=1.2, 1.5, 2$ and $3$ with the set of
parameters $(\lmdb,M)=(1,0.9)$, corresponding to a contracting map.
Starting from $b\sbkn\spa$ on the diagonal line, the dynamic forecast
takes  $b\sbkn\spa$ to $\bk\spf$  on the corresponding $\mu$-curve
(upper group), which is followed by the probabilistic analysis down
to $\bk\spa$ on the corresponding  $\mu$-curve (lower group) to
complete one data  assimilation cycle, $b\sbkn\spa\to\bk\spf\to\bk\spa$.
A new analysis $\bk\spa$ is moved horizontally onto the diagonal
line to become an initial scale factor value for the next cycle.

On each $\bk\spa$ curve for a fixed $\mu$, the symbols
(circle, square, diamond and triangle for
$\mu=1.2, 1.5, 2$ and $3$, respectively) at the intersection with the
diagonal line, i.e. $\bb\spa=\bk\spa=b\sbkn\spa$, are the stable
solutions of the KL filter.
The maps (\ref{jfjnnkkk}) and (\ref{jajnnkkk}) in fact have each a
stable fixed-point solution, $\bb\spf$ and $\bb\spa$, that attracts any initial
condition for any exponent $\mu$, given  a set of parameters $(\lmdb, M)$.
For $\mu=2$ retrieving the case of the KG filter, this stable fixed-point
can be
obtained analytically\cite{GCTBI}.

The error probability distributions for the stable fixed-points corresponding
to the Student's distribution (\ref{fdanhfal})  are shown
in Figure \ref{fig:unimap}b--e based on the scale factor $\bb\spfa$.
We use $(x\spf,y\spo)=(0,1)$ and $H=1$, so that $x\spa=K^{L}$.
For small $\mu$, the KL filter favors more strongly the better sample
characterized by
the smaller scale factor between the two ($\bb\spf$ for $x\spf$ and
$(\lmdb)^{\mu}$
for $y\spo$). This effect is stronger when $\mu$ decreases to $1$, as
discussed in
section 2.
The value of the fixed point $\bb\spa$ is larger for smaller $\mu$, i.e.,
a system with a probability distribution with heavier tail has
greater uncertainty, not only because of its slow decay measured by the
exponent $\mu$ but also due to its overall amplitude quantified by the
scale factor.
Furthermore, the slope of the curve $\bk\spa$ as a function of $b_{k-1}^a$
shown in Figure \ref{fig:unimap}
is closer to the horizontal for smaller $\mu$, indicating that the convergence
to the stable fixed-point is faster for the heavy-tail probability
distribution.
This is because the KL filter with smaller $\mu$ tends to favor
either forecast or observation strongly, depending on their relative noise
amplitude quantified by the scale factor of their noises,
as discussed in Section \ref{sec:uni-prop}.

\begin{figure}
\begin{center}
\epsfig{file=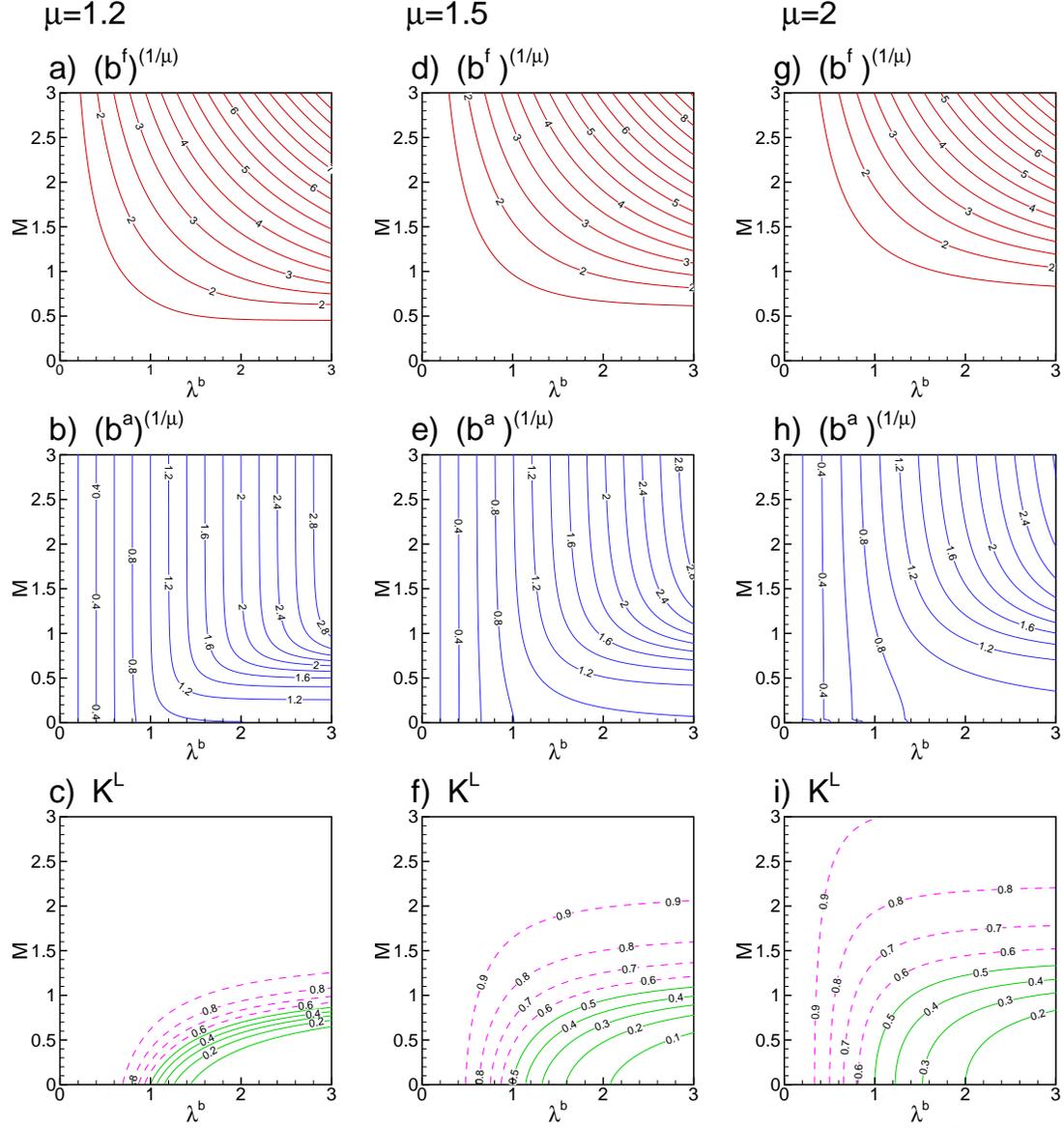,height=16cm}
\caption{\protect\label{fig:KLMbH} Stable fixed-point of the
Kalman-L\'{e}vy filter as a function of $\lmdb$ and $M$:
a) $(\bb\spf)^{{1\over \mu}}$, b) $(\bb\spa)^{{1\over \mu}}$ and
	c) $\KbL$ for $\mu=1.2$;
d), e) and f) same as a), b) and c) but for for $\mu=1.5$;
g), h) and i) same as a), b) and c) but for for $\mu=2$.}
\end{center}
\end{figure}

Since the stationary KL assimilation system quickly
approaches a unique steady state for a given set of parameters
$(\lmdb,M)$ and exponent $\mu$, the stable fixed-point defined
by $\bb\spf$ and $\bb\spa$ along with the corresponding optimal
Kalman-L\'{e}vy gain $\KbL$ suffice to provide a complete description
of the stationary assimilation process.
Figure \ref{fig:KLMbH} shows the stable solution obtained from
(\ref{jfjnnkkk}), (\ref{jajnnkkk}) and (\ref{eq:uni-KL}), which
are plotted in the parameter space $(\lmdb,M)$ for
$\mu=1.2, 1.5$ and $2$.
All tail covariances (scale factors) are plotted in terms
of $(\bb)^{{1 \over \mu}}$ so as to preserve the characteristic
scale as in (\ref{jnadsfajkaka}) of the error, independently
of the value of $\mu$.

For convergent dynamics  $M<1$ with sufficiently large
relative observational error $\lmdb>1$,
$\lambda^H\gg 1$ and hence $\KbL\approx 0$ favors the forecast
(Figure \ref{fig:KLMbH}c, f and i) and hence results in
$\bb\spa \sim \bb\spf$.
For divergent dynamics with a larger value of $M>1$,
however, it yields a large value for the forecast's scale factor
$\bb\spf$ (Figure \ref{fig:KLMbH}a, d and g) with respect to
relative observational error $\lmdb$. Here, the KL filter correctly
derives that the errors are amplified by the unstable evolution of
the system. Since $\bb\spf$ is large and hence $\lambda^H\approx 0$,
$\KbL\approx 1$ favors the observation over the forecast and hence
results in  $\bb\spa \sim (\lmdb)^{\mu}$
(Figure \ref{fig:KLMbH}b, e and h).
This effect derived by the KL filter is more
significant for heavy-tail probability distributions with a
smaller $\mu$, and it is manifested in the much steeper gradient
of $\KbL$ for $\mu=1.2$ (Figure \ref{fig:KLMbH}c) in contrast
to the widely spread contour maps observed for $\mu=2$
(Figure \ref{fig:KLMbH}i).

Note that for $M=0$, we retrieve the univariate
filter studied in section 2. In particular, for $\lmdb=1$,
$K^{L}=1/2$, corresponding to equal weights of the assimilation
on observation and forecast for any exponent $\mu$.
The curvature to the right taken by the contour maps of the Kalman gain $K$
as a function of
$M$ can easily be rationalized: a larger $M$ implies a larger forecast error,
hence a smaller effective $\lmdb$. One thus needs a large observation over
dynamical
error ratio to get the same effective effect, hence the downward convexity
of the
contour maps.

\subsubsection{Relative performance of the KL and KG filters}
\label{sec:KLF-uni-KLG}

We now examine the case where the KG filter $K^{G}$ (corresponding
to putting $\mu=2$ in the solutions) is
applied to the system whose true noise distribution is the heavy-tail
power law ($\mu<2$). This may happen in a practical implementation
of Kalman filtering
when we do not know the nature of the noises very well and a finite variance
assumption is made. This is also probably the only choice left to the 
operator in
absence of our solution presented in this paper. This exercise is thus
aimed at quantifying what we have gained concretely by recognizing the
non-Gaussian nature of the noise and by providing the corresponding solution.
We stress that what matters according to the KG framework is whether 
the noise has
a finite variance or not. In other words, all non-Gaussian noises
with finite variance are treated in the same fashion within the KG
approach by analyzing the variance only. In contrast, the KL solution
distinguishes noises even if they have a finite variance by analyzing 
the structure
of their `fat-tail'' characterized by the exponent $\mu$. For instance,
the KL gain is different for noise distributions with power law
tails with $\mu=3$ and with $\mu=4$, while in contrast the KG solution
is the same for both if they have the same variance.

We formulate this scenario in a general form when an incorrect
model exponent $\mum$ is used to assimilate the data from the system
with true exponent $\mu$.
In case of the KG filter application, this means $\mum=2$.
The model for data assimilation that we obtain is equivalent to
(\ref{eq:uni-xkf})--(\ref{eq:uni-lambda}) by replacing
\ba
(x\sbk\spfa,B\sbk\spfa, K\sbk) & \longrightarrow &
   (\xm\sbk\spfa,\Bm\sbk\spfa, \Km\sbk)~,
  \label{eq:uni-modelxB} \\
  \left[ \mu, B\sbk^{\eta, \epsilon}, H\lambda\sbk^{H} \right]
  & \longrightarrow &
[ \mum,(\Bm\sbk^{\eta, \epsilon})^{{\mu\over \mum}},
(H\mdl{\lambda}\sbk^{H})^{{\mu\over \mum}} ]~, \label{eq:uni-modelp}
\ea
where $\mdl{\{\cdot\}}$ represents the model filtering.
The exponent factor $\{\cdot\}^{{\mu\over \mum}}$ arises so as to
preserve the characteristic scale of the noises.

Use of the model gain $\Km\sbk\neq K\sbk^{L}$ due to an incorrect
model exponent $\mum$ yields a non-optimal filtering by definition,
in a sense that the analysis scale factor $B\sbk\spa$ is not a minimum.
In addition, such a model filtering estimates both forecast
and analysis tail covariances $\Bh\sbk\spfa$ incorrectly as
$\Bm\sbk\spfa$,  because the real evolution of the tail covariances using the
non-optimal model gain $\Km\sbk$ should follow the non-optimal
KL filtering scheme which itself uses the true exponent $\mu$
\ba
\Bh\sbk\spf & = & |\Mkkn|^{\mu} ~ \Bh\sbkn\spa + B\sbkn\spet~,
	\label{eq:uni-Bhf} \\
\Bh\sbk\spa & = & |1-\Km\sbk H\sbk|^{\mu} ~ \Bh\sbk\spf
    + (\Km\sbk)^{\mu} ~ B\sbk\spep~.
	\label{eq:uni-Bha}
\ea

Accordingly, there are three filtering representations, i.e.,
\begin{itemize}
\item[(i)]  true and optimal KL filtering $B\sbk\spfa$ using $(\mu,K^{L})$;
\item[(ii)]  true but non-optimal filtering $\Bh\sbk\spfa$ using
   $(\mu,\Km^{G})$;
\item[(iii)]  model and  incorrect filtering $\Bm\sbk\spfa$  using
  $(\mum,\Km^{G})$.
\end{itemize}
By definition, the optimal filtering (i) is always superior to
the non-optimal filtering (ii),
$(B\sbk\spa)^{1\over\mu}\leq (\Bh\sbk\spa)^{1\over\mu}$.
It is possible, however, that the incorrect model filtering (iii) returns
a value for the scale factor which is numerically smaller (see table 1).
Since the
model exponent $\mum$ is different from the true exponent $\mu$, the
scale factors cannot be compared directly to infer the quality of the
assimilation process.

When the system is time-independent, the normalized one-dimensional maps
of the tail covariances using the non-optimal KL filter with model gain
$\Km$ are
\ba
\bh\sbk\spf & = & |M(1-\Km  H)|^{\mu} ~ \bh\sbkn\spf
   + |M \Km  H \lmdb|^{\mu} + 1 ~,
	\label{eq:uni-bhf} \\
\bh\sbk\spa & = & |M(1-\Km  H)|^{\mu} ~  \bh\sbkn\spa
   + |1-\Km  H|^{\mu} + (\Km  H \lmdb)^{\mu}~.
	\label{eq:uni-bha}
\ea
This non-optimal filtering also has a unique stable fixed-point
\ba
{\bar \bh}\spf & = & {|M \Km  H \lmdb|^{\mu} + 1 \over
  1- |M(1-\Km  H)|^{\mu}}~, \label{eq:uni-bhfb} \\
{\bar \bh}\spa & = & {|1-\Km  H|^{\mu} + (\Km  H \lmdb)^{\mu}~ \over
  1- |M(1-\Km  H)|^{\mu}}~, \label{eq:uni-bhab}
\ea
provided the condition for stability is satisfied
\be
  0<|M(1-\Km  H)|<1~. \label{eq:uni-bhstbl}
\ee

\begin{figure}
\begin{center}
\epsfig{file=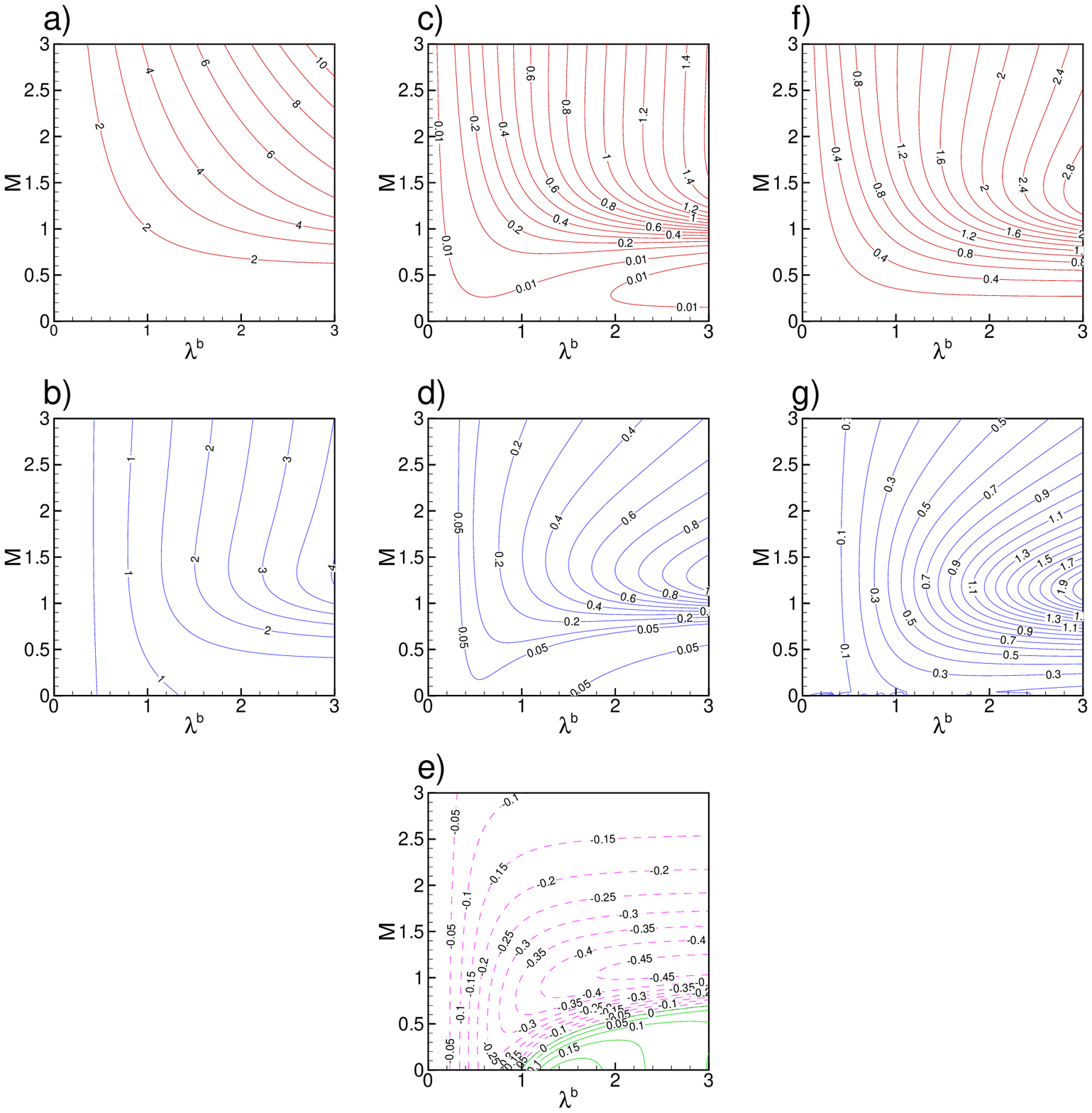,height=16cm}
\caption{\protect\label{fig:KGMbH} Stable fixed-point of the
Kalman-Gaussian filter applied to a heavy-tail system with
$\mu=1.2$: a) $(\bh\spf)^{{1\over\mu}}$, b) $(\bh\spa)^{{1\over\mu}}$,
c) difference
    $({\bar \bh}\spf)^{{1\over\mu}}-(\bb\spf)^{{1\over\mu}}$
   between the non-optimal filtering solution
    $({\bar \bh}\spf)^{{1\over\mu}}$ shown in a)
    and the optimal KL filtering solution $(\bb\spf)^{{1\over\mu}}$
    shown in  Figure \ref{fig:KLMbH}a,
d) difference $({\bh}\spa)^{{1\over\mu}}-(\bb\spa)^{{1\over\mu}}$
    between the non-optimal filtering solution
   $({\bh}\spa)^{{1\over\mu}}$ shown in b) and the
   optimal KL filtering solution $(\bb\spa)^{{1\over\mu}}$ shown in
    Figure \ref{fig:KLMbH}b,
e) difference $K^{G}-K^{L}$ between the non-optimal filtering solution
    and the model filtering solution.
f) difference
    $({\bh}\spf)^{{1\over\mu}}-(\bm\spf)^{{1\over\mum}}$
   between the
   non-optimal filtering solution $({\bh}\spf)^{{1\over\mu}}$
   shown in b) and the model filtering solution
   $({\bm}\spf)^{{1\over\mum}}$,
g) difference
    $({\bh}\spa)^{{1\over\mu}}-(\bm\spa)^{{1\over\mum}}$
   between the
   non-optimal filtering solution $({\bh}\spa)^{{1\over\mu}}$
   shown in b) and the mode filtering solution
   $({\bm}\spa)^{{1\over\mum}}$.}
\end{center}
\end{figure}

To see this effect  of non-optimal filtering for the KG filter
application,  we apply the model gain $\Km=\Km^{G}$ with $\mu=2$
(Figure \ref{fig:KLMbH}i) to a time independent system
(\ref{jfjnnkkk})--(\ref{eq:uni-KL}) subjected to the L\'evy noise
with true exponent $\mu=1.2$.
The stable assimilation cycle for this optimal KL filtering
have been presented  in Figure \ref{fig:KLMbH}a--c.
The unique stable fixed-points of the non-optimal filter given by
(\ref{eq:uni-bhfb}) and  (\ref{eq:uni-bhab}) are shown in
Figures \ref{fig:KGMbH}a and b,
  in terms of the characteristic scale $({\bar \bh}\spfa)^{1\over \mu}$.
Because the KG filtering is no longer optimal,
${\bar \bh}\spfa$ are now larger than the corresponding optimal
scale factors $\bb\spfa$ of the KL fixed-point  (Figure \ref{fig:KLMbH}a
and b).

To quantify the difference between the KL and KG solutions,
we construct the differences of the  normalized stable fixed-point
found in the three assimilation representations (i)--(iii).
In Figure \ref{fig:KGMbH}c--f, we present the comparison for the
following two cases:
\begin{enumerate}
\item difference between the non-optimal filtering
(ii: as in Figure  \ref{fig:KGMbH}a and b) and
optimal KL filtering (i: as in Figure \ref{fig:KLMbH}a and b);
\item difference between non-optimal filtering (ii)
and incorrect model filtering (iii).
\end{enumerate}
All results are shown in terms of the characteristic error scales,
$b^{1\over\mu}$ or $b^{1\over\mum}$,
so that the comparison can be made independently of the exponents
$\mu$ and $\mum$ in the filters.

The first comparison between (ii) and (i)
corresponds to the difference between the optimal
and non-optimal filtering.
In Figure \ref{fig:KGMbH}c and d, we observe a bimodal structure in
the difference
$({\bh}\spfa)^{{1\over\mu}} - (\bb\spfa)^{{1\over\mu}}$,
caused by the maximum-minimum structure in the gain
$\Km^{G} - \KbL$ (Figure \ref{fig:KGMbH}e), whose
origin is the following.
For $M<1$ for which $M^{2}<M^{\mu}$ for $\mu<2$, $\Km^{G} >\KbL$.
The non-optimal gain $\Km^{G}$ obtained from the model KG solution
thus overestimates the uncertainty of  the forecast.
On the other hand, for $M>1$ for which $M^{2}>M^{\mu}$ for $\mu<2$,
$\Km^{G}<\KbL$.
The non-optimal gain $\Km^{G}$ now underestimates the reliability
of the observation.
Both ways, it increases the  non-optimal solution ${\bh}\spa$ in
comparison  to the optimal solution $\bb\spa$.

The second comparison between (ii) and (iii) relates to
the  actual mistake in the  model solution ${\bm}\spfa$ made
  by using the model gain $\Km^{G}$ to
the system which reaches a different, non-optimal stable fixed-point
${\bh}\spfa$.
As shown in \ref{fig:KGMbH}f and g,
$({\bh}\spfa)^{{1\over\mu} }- ({\bm}\spfa)^{{1\over\mum}}$
are positive and therefore
the model filtering incorrectly underestimates the error.

\begin{figure}
\begin{center}
\epsfig{file=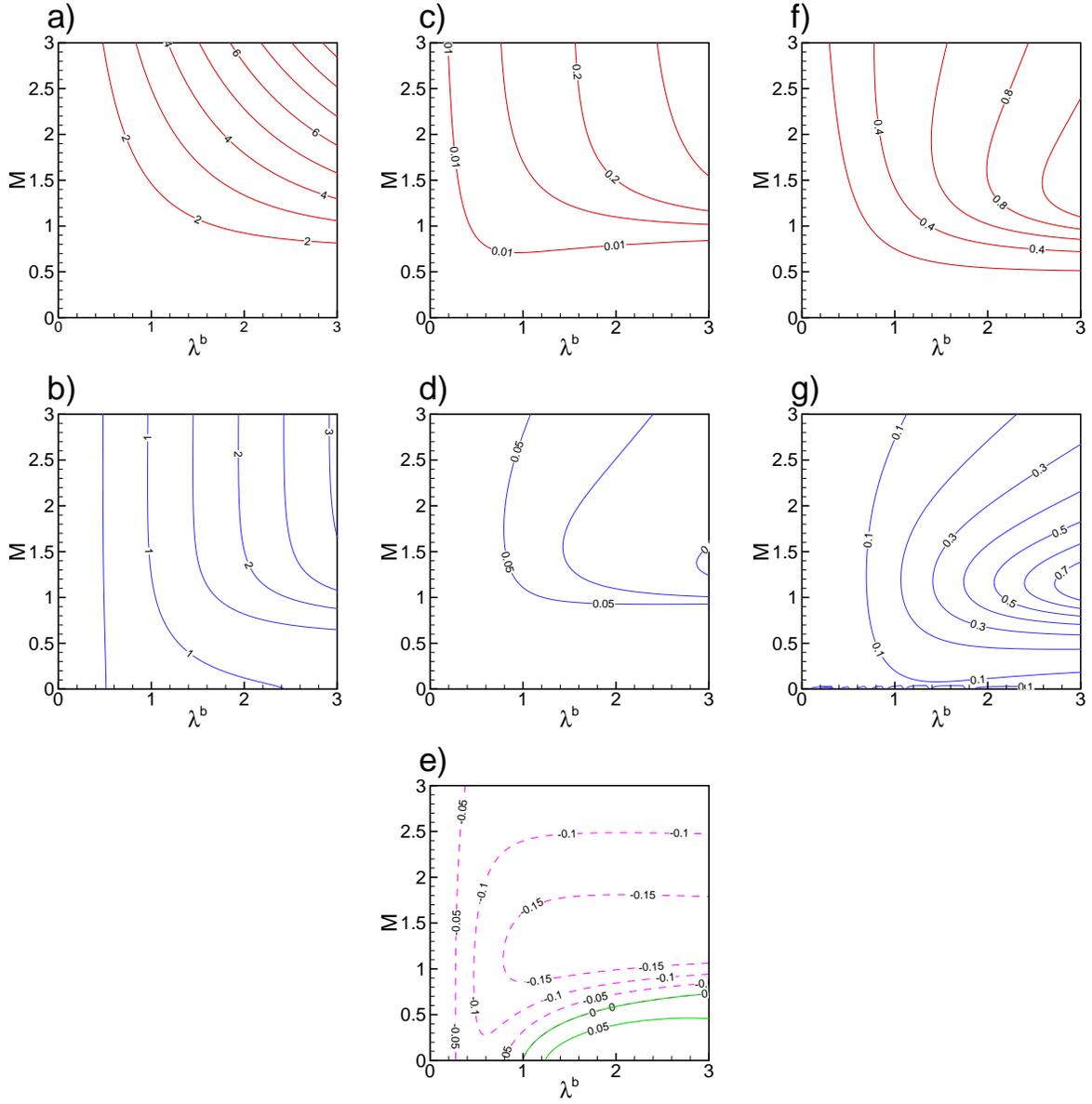,height=16cm}
\caption{\protect\label{fig1.5:KGMbH} Same as figure \ref{fig:KGMbH}
for dynamical and observational noises given by L\'evy distributions
with exponent $\mu=1.5$. }
\end{center}
\end{figure}

Figure \ref{fig1.5:KGMbH} is the KG filtering application when
the true exponent $\mu=1.5$ is not as heavy as the previous
case $\mu=1.2$.
Although the KL solution is better than the KG one as expected,
the difference is smaller: the improvement is of the order
of $5-10\%$ at most. The fact that the improvement has a smaller amplitude
is clear: a larger exponent implies a thinner tail and thus a behavior
closer to the Gaussian case.
Recall that at $\mu$ goes to $2$, the Gaussian
case and solution are recovered.

\subsubsection{Numerical experiment}
\label{sec:KLF-uni-exp}

To check these results derived from the analysis of the deterministic
behavior of the
tail covariances (scale factors) $\bk\spfa$, we present a numerical
experiment of the stochastic dynamics, observation construction and
assimilation processes.
We use the parameter set $(\lmdb,M)=(1,0.9)$  with noises distributed
according to  a L\'evy law with exponent $\mu=1.2$.
The one-dimensional map and probability distribution of the
stable fixed-point for this parameter set are shown in
Figure \ref{fig:unimap}a and b.
To generate the L\'evy noises, we follow the standard algorithm
described initially in \cite{Chambers} and use the software
available at \cite{McCulloch}.
The stochastic variables $x\sbk\spt$ and $y\sbk\spo$ are
generated over $10000$ time steps.

\begin{figure}
\begin{center}
\epsfig{file=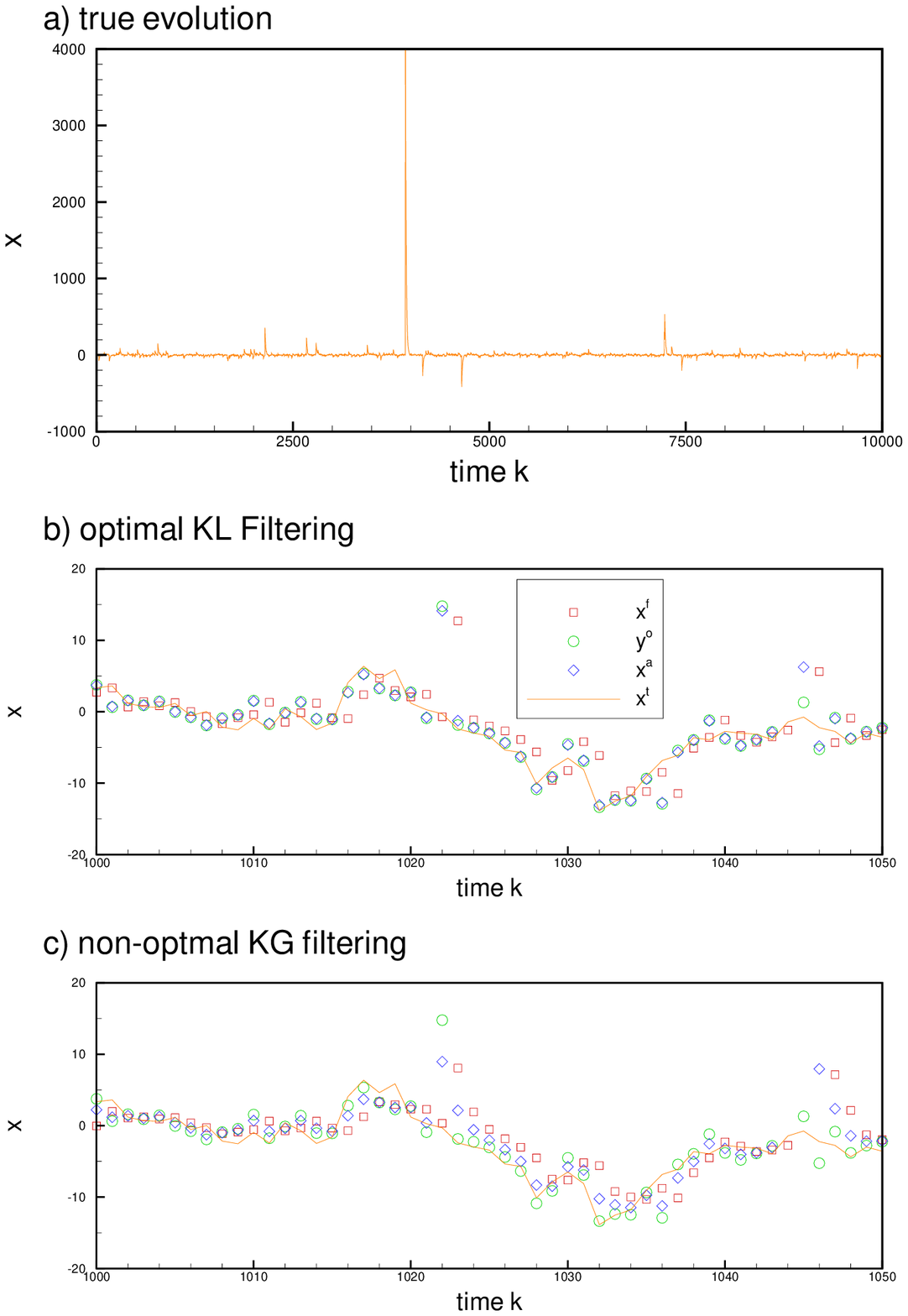,height=16cm}
\caption{\protect\label{fig:kmu12}
Numerical experiment for the parameter set $(\lmdb,M)=(1,0.9)$
with $\mu=1.2$; a) evolution of true state variable $x\spt\sbk$
for $k=1,\ldots,10000$. Note the occurrence of a few large peaks corresponding
to rare but extreme noise fluctuations distributed with the L\'evy
distribution.
Panels b) and c) show a magnification of panel a) in the time interval
$[1000, 1050]$
and compare this true dynamics (continuous line)
  with the forecasts $x^f$ (squares), the observations
$y^o$ (circles) and the assimilation analysis $x^a$ (diamonds).
Panel b) corresponds to the use of the
optimal KL filtering while panel c) corresponds to the
the model filtering with $\mum=2$, i.e. standard Kalman-Gaussian filter.
It appears clear by visual inspection that the optimal KL analysis $x^a$ is
much closer
more often than not to the true dynamics.}
\end{center}
\end{figure}

Typical results of the KL filtering $(\mu,K^{k}\sbk)$
are shown in Figure \ref{fig:kmu12}.
The true evolution $x\spt\sbk$ is shown over the $10000$ time steps in
Figure \ref{fig:kmu12}a. Note the occurrence of a few very large fluctuations
that dwarf most of the remaining dynamics. To get a closer view, we enlarge
figure \ref{fig:kmu12}a in the narrow time interval $[1000, 1050]$ shown in
figures
\ref{fig:kmu12}b-c. Figure \ref{fig:kmu12}b gives the dynamical evolution
of the true,
forecasted, observation and analysis variables, when using
the optimal KL filtering, while figure \ref{fig:kmu12}c corresponds to the
use of the non-optimal filtering.
Note that the two filtering's use the same gain $\Km^{G}$
and therefore result in the same $\xm\spfa$.
One can observe on Figure \ref{fig:kmu12}b that the
optimal KL filtering $x\spa\sbk$ follows rather closely the observation
$y\spo\sbk$. This results from
the high value of $\KbL$. In constrast,
the non-optimal filtering puts $x\spa\sbk$ midway between
$y\spo\sbk$ and $x\spf\sbk$.
The tail covariances (scale factor) quickly approaches
the stable fixed-point after a few iteration
as given in Table \ref{tbl:bK},
along with the stable fixed-points of the non-optimal filtering
${\bh}\spfa$ with $(\mu,\Km^{G}\sbk)$ and
model filtering ${\bm}\spfa$ with $(\mum,\Km^{G}\sbk)$.

\begin{table}[!ht]
\begin{center}
\begin{tabular}{|c||c|c|c|} \hline
  & \makebox[4.cm][c]{(i) optimal KL \ $(\bb)$}
  & \makebox[4.cm][c]{(ii) non-optimal \ $({\bar \bh})$}
  & \makebox[4.cm][c]{(iii) model \ $({\bar \bm})$} \\
  &  $(\mu,\KbL)$  &  $(\mu,\Km^{G})$  &  $(\mum,\Km^{G})$ \\ \hline \hline
$b\spf$ & 1.87 & 2.09 & 1.48  \\ \hline
$b\spa$ & 0.99 & 1.24 & 0.59  \\ \hline
$K$ & 0.96 & \multicolumn{2}{c|}{0.60}  \\ \hline
\end{tabular}
\caption{Stable fixed-points of (i) optimal KL, (ii) non-optimal KG and
(iii) incorrect KG, using the exponents $(\mu,\mum)=(1.2,2)$ and
system parameter set $(\lmdb,M)=(1,0.9)$.}
\label{tbl:bK}
\end{center}
\end{table}

Because the KL filter is designed for the global control of the uncertainty
by minimizing the tail covariance, we propose to compare the tails
of the distribution of errors $(x\spa-x\spt)$ resulting from the
two methods (KL and KG) to assess their relative performance.
Note that, since
the covariance does not exist for $\mu<2$, it cannot be
used to evaluate the performance of the
heavy-tail KL filtering.

\begin{figure}
\begin{center}
\epsfig{file=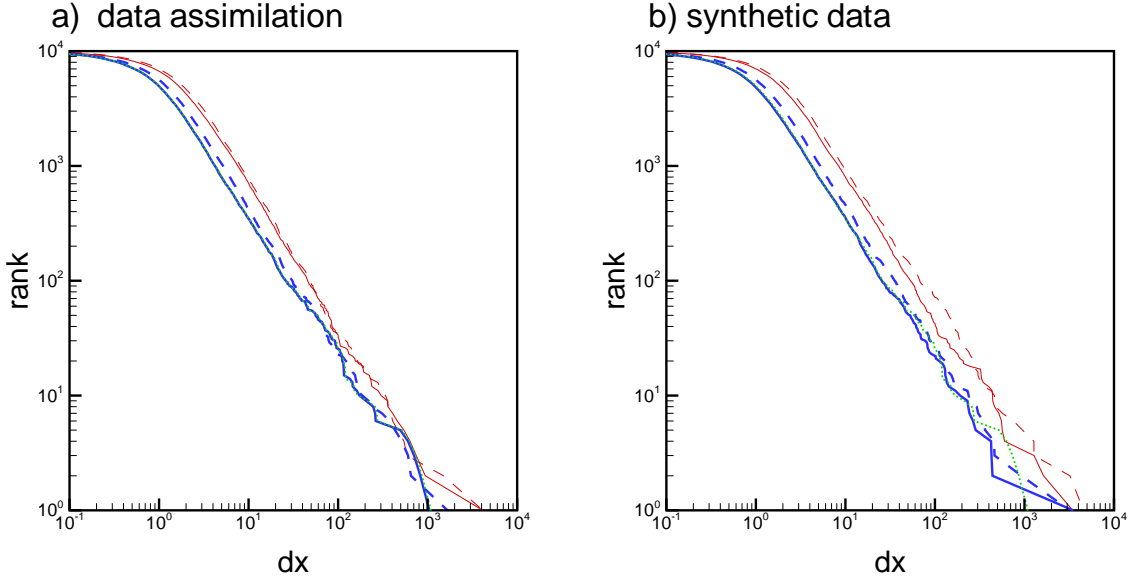,height=8cm}
\caption{\protect\label{fig:rnk}
a) (Complementary) cumulative distribution
(i.e. number of time steps where the error is larger than a value
read on the abscissa) of the error  between the analysis and
the true trajectory
for $(\lmdb,M)=(1,0.9)$ and $\mu=1.2$, obtained by using the
optimal KL filter $K^{L}$ ($x\spa-x\spt$; in thick solid line) and
model KG filter $\Km\sbk$  ($\xm\spa-x\spt$; in thick dashed line). We observe
clearly that the distribution of $x\spa-x\spt$ is below that of
$\xm\spa-x\spt$,
i.e. the errors are globally reduced {\it in distribution} by
application of the Kalman-L\'evy method compared to the standard
Kalman-Gaussian method.
We also show the cumulative distributions of
the difference between forecast and true trajectory for the
optimal KL filter $K^{L}$ ($x\spf-x\spt$; in solid line) and
model KG filter $\Km\sbk$  ($\xm\spf-x\spt$; in dashed line),
as well as the cumulative distributions of the observations
($y\spo/H-x\spt$; in dotted line, which is almost identical to
$x\spa-x\spt$ and thus hardly visible due to the thickness of the lines).
b) (Complementary) cumulative distribution of the synthetic
simulation of random L\'{e}vy variables based on the
scale-factors of the stable fixed-points $\bb\spfa$ and ${\bar \bh}\spfa$
(Table \ref{tbl:bK}) predicted by the theory.
The cumulative distributions of the same observations
($y\spo/H-x\spt$; in dotted line) as in a) is also shown for reference.
This alternative method for constructive the distributions of errors shows
the full consistency of the approach.}
\end{center}
\end{figure}

Figure \ref{fig:rnk}a shows the (complementary) cumulative
distribution of $(x\spa-x\spt)$
and $(x\spf-x\spt)$, as well as that of $(y\spo/H)- x\spt$ for reference.
For this parameter set $(\lmdb,M)=(1,0.9)$, the two optimal KL and
model KG gains at their stable fixed-points differ by 37.5\% (Table
\ref{tbl:bK}).
This shows that the model KG filter underestimates the reliability of
observation and overestimates the value of the forecast.

Although the difference in the cumulative distributions is rather
subtle to determine from visual inspection of Figure \ref{fig:rnk}a,
the cumulative distribution of the error between
the analysis and the true trajectory obtained from the optimal
KL filter is consistently below that obtained by using the non-optimal
filter, apart from expected fluctuations.
In probability terms, the optimal KL error distribution
exhibit the property of being ``stochastically dominant'' over the
model KG error distribution.
This shows that the optimal KL filter is indeed superior to the
model KG filter in the presence of heavy tails.

In fact, our theory predict the difference of 37.5\% (Table \ref{tbl:bK})
based on the stable fixed-point presented in Section \ref{sec:KLF-uni-prop}.
It is confirmed by the synthetic simulation of the L\'{e}vy random
variables with $\mu=1.2$ as presented in Figure \ref{fig:rnk}b
based on the corresponding scale factors of the stable fixed-point
$\bb\spfa$ and ${\bh}\spfa$ (Table \ref{tbl:bK}).

The superficial visual effect in Figure \ref{fig:rnk}a
can be explained as follows.
In the tails, L\'evy laws with exponent $\mu$ are power law
given by $C/(x\spa-x\spt)^{\mu}$ where $C$ is the scale factor
of the errors.
If $C$ is higher by $37.5\%$ for the non-optimal filtering compared to
the optimal KL filtering (Table \ref{tbl:bK}),
this represents a significant error reduction.
However, this will not be strikingly
visible in the log-log representation of figure \ref{fig:rnk}, since
$\ln 1.375 \approx 0.32$ leads to a translation of the two cumulative
distributions by only $0.32$, hence the small but still visible effect.

Another more compact way of quantifying the relative performance of the two
solutions is to calculate a typical error amplitude, which generalizes the
covariance. Since we have considered the situation where $\mu > 1$, the
average of the {\it absolute value} $\langle |x\spa-x\spt| \rangle$
  of the errors corresponds to a moment of
order $1$, which is defined mathematically and is numerically well-behaved.
Our direct numerical simulations show a decrease of the typical error
amplitude ($\langle |x\spa-x\spt| \rangle$) by approximately $20\%$
when going from the non-optimal solution ($\approx 3.3$) to the optimal
KL filtering ($\approx 2.8$).

\section{Conclusion}

We have presented the solution of the Kalman filter problem for
dynamical and forecast noises distributed according to power laws
and L\'evy laws. The main theoretical concept that we have used is
to optimize the Kalman filter to chisel the tail of the distribution of
residual errors so as to minimize it globally. In order to implement this
program, we have introduced the concept of a ``tail covariance'' that
generalizes the usual notion of the covariance. The full solution, called
the Kalman-L\'evy filter, is obtained by the solution of a general non-linear
equation. We have investigated in detail the quality of this solution in
the univariate case and have shown by direct numerical experiments
that the improvement is significant, all
the more so, the heavier the tail, i.e. the smaller the power law exponent
$\mu$.

\newpage
\setcounter{section}{0}
\renewcommand{\thesection}{\Alph{section}}

\section{Stable L\'{e}vy laws}
\label{Appendix:levy}

The stable laws have been studied and classified by Paul L\'evy, who
discovered that, in addition to the Gaussian law, there is a large number
of other pdf's sharing the stability condition
\be
P_N(x') dx' = P_1(x) dx  ~~~~~\mbox{  where } ~~~x'=a_N x+b_N ~,
\label{stablelfdf}
\ee
for some constants $a_N$ and $b_N$, where $x'$ is the sum of $N$
independent variables
of the type $x$
distributed according to the pdf $P_1(x)$.
One of their most interesting properties is their asymptotic power law
behavior.

A symmetric L\'evy law centered on zero is completely characterized
by two parameters which can be extracted solely from its asymptotic dependence
\be\label{E_ASYM2}
P(x) \sim  \frac{C}{|x|^{1+\mu}} ~~~~~~\mbox{ for $x\rightarrow\pm\infty$}~ .
\ee
  $C$ is a positive constant called the tail or scale parameter and the exponent
$\mu$ is between $0$ and $2$ ($0<\mu< 2$). Clearly, $\mu>0$  for the pdf to be
normalizable. As for the other condition $\mu< 2$, a pdf with a
power law tail with $\mu > 2$ has a finite variance and thus converges
(slowly) in
probability to the Gaussian law. It is therefore not stable. Only its
shrinking tail for
$\mu > 2$ remains of the power law form. In contrast, the whole L\'evy
pdf remains stable for $\mu< 2$.

All symmetric L\'evy law with the same exponent $\mu$ can be obtained from the
L\'evy law $L_{\mu}(x)$ with the same exponent $\mu$, centered on zero and with
unit scale parameter $C=1$, under translation and rescaling transformations
\be
P(x)dx=L_{\mu}(x')dx' ~~~~~ \mbox{ where } x'=C^{1/\mu}x+ m~,
\label{E_CHANGE_LEVY}
\ee
$m$ being the center parameter.

L\'evy laws can be asymmetric and the parameter quantifying
this asymmetry is
\be
\beta=(C_{+}-C_{-})/(C_{+}+C_{-})~,
\ee
where
$C_{\pm}$ are the scale parameters for the asymptotic behavior of the L\'evy
law for $x \to \pm \infty$. When $\beta\neq 0$, one defines a unique scale
parameter $C=(C_{+}+C_{-})/2$, which together with $\beta$ allows one to
describe the behavior at $x \to \pm \infty$. The completely antisymmetric case
$\beta=+ 1$ (resp. $-1$) corresponds to the maximum asymmetry.

For $0<\mu<1$ and $\beta = \pm 1$,
the random variables take only positive (resp. negative) values.

For $1<\mu<2$ and $\beta = +1$,
  the L\'evy law is a power law for $x \to +\infty$ but goes to zero
for $x \to -\infty$ as $P(x) \sim \exp(-|x|^{\mu/\mu-1})$ .
This decay is faster than the Gaussian law. The symmetric situation is found
for $\beta = -1$.

An important consequence of (\ref{E_ASYM2}) is that the variance of a
L\'evy law
is infinite as the pdf does not decay sufficiently rapidly at $|x| \to \infty$.
  When $\mu\leq 1$, the L\'evy law
decays so slowly that even the mean and the average of the absolute
value of the spread diverge. The median and the most probable value
still exist and coincide, for symmetric pdf ($\beta = 0$), with the center $m$.
The characteristic scales of the fluctuations are determined by the scale
parameter $C$, {\it i.e.} they are of the order of $C^{1/\mu}$.

There are no simple analytic expression of the symmetric L\'evy stable laws
  $L_{\mu}(x)$, except for a few special cases. The best known is
  $\mu=1$, called the Cauchy\index{Cauchy distribution} (or Lorentz) law,
\be
L_{1}(x)=\frac{1}{x^{2}+\pi^{2}}~~~~~{\rm for}~-\infty < x < +\infty~.
\ee
The L\'evy law for $\mu = 1/2$ is \cite{Montrollwest}
\be
L_{1/2}(x) = {2 \over \sqrt{\pi}}~{\exp \left(-{1 \over 2x}\right) \over
(2x)^{3 \over 2}}~~~~
{\rm for}~x>0~.
\ee
This pdf $L_{1/2}(x)$ gives the distribution of first returns to origin
of an unbiased random walk.

L\'evy laws are fully characterized by the expression of their
characteristic functions\,:
\be
\hat{L}_{\mu}(k)=\exp\left(-a_{\mu}|k|^{\mu}\right)
\ee
where $a_{\mu}$ is a constant proportional to the scale parameter
$C$\,:
\be
a_\mu=\frac{\pi~C}{\mu^2\Gamma(\mu-1)\sin\left(\frac{\pi\mu}{2}\right)} ~~~~~
  ~~~~~ \mbox{      for } ~~ 1 < \mu < 2 .
\ee
A similar expression holds for $0 < \mu < 1$, while $\mu = 1$ and $2$ requires
a special form (see \cite{GK} for full details).
For $\beta\neq 0$, we have
\be
\hat{L}_{\mu}^{\beta}(k)=
\exp\left[-a_{\mu}|k|^{\mu}\left(1+i\beta\tan(\mu\pi/2)
\frac{k}{|k|}\right)\right]  \mbox{   for } \mu \neq 1 ~.
\label{fhjuixw}
\ee
For $\mu = 1$, $\tan(\mu\pi/2)$ is replaced by $(2/\pi) \ln |k|$.

\newpage

\section{Proof of the optimality of the solution of (44)
}
\label{issues}

To prove the optimality of the solution $\mK\spo$ solving (\ref{eq:dtrBdAij}),
it is sufficient to consider only one of the system of $N$ equations 
for a single
and fixed index $i$. We thus drop the index $i$ and
  define the matrix $\mtrx{\Omega}\spfo$ in
$\bbR^{N\times N}$ and the vector $\vctrg{\kappa}$ in $\bbR^{N}$ as follows:
\ba\label{eq:Ok}
\Omega\spf_{pq}&=&\left\{ \begin{array}{ll}
\left|G\spf_{ip}- \sum_{m=1}^{N}K\spo_{im}G\spf_{mp}\right|^{\mu-2}
  C\spf_{p} & \mbox{for $p=q$} \\ 0, & \mbox{for $p\ne q$}
	\end{array} \right. \nonumber \\
\Omega\spo_{pq}&=&\left\{ \begin{array}{ll}
  \left|\sum_{m=1}^{N}K\spo_{im}G\spo_{mp}\right|^{\mu-2}
  C\spo_{p} & \mbox{for $p=q$} \\ 0, & \mbox{for $p\ne q$}
	\end{array} \right. \nonumber \\
\kappa_{q}&=&K\spo_{iq}~~.
\ea
Expression (\ref{eq:ddtrBddAij}) can then be written as:
\ba\label{eq:newdef}
\pdrv{^{2}}{\vctrg{\kappa}^{2}} {\hat B}_{ii}
  &=& \mu (\mu-1)  \left[
  \mG\spf {\bf \Omega}\spf ( \mG\spf ) \spT +
  \mG\spo {\bf \Omega}\spo ( \mG\spo ) \spT \right]~.
\ea
It is clear in this form that the Hessian matrix is positive definite for
$\mu>1$ because the linear sum of symmetric positive definite matrices
results in a positive definite matrix.

There are two issues associated with this formulation for the
optimal $\vctrg{\kappa}$ (i.e., $\mK\spo$): 1) the possible existence of
singular terms for $\mu<2$;
and 2) the solvability of the nonlinear system.
To address these issues, we rewrite (\ref{eq:dtrBdAij}) as
\ba\label{eq:fKmu}
  \pdrv{}{\vctrg{\kappa}} \trace \ {\hat B}_{ii}
  = \vctr{f}(\vctrg{\kappa},\mu)=0,
\ea
and seek for a solution branch $\vctrg{\kappa}(\mu)$
of (\ref{eq:fKmu}) using implicit function theory as $\mu$ varies.

At $\mu=2$, the system is linear in $\vctrg{\kappa}$
and a unique solution $\vctrg{\kappa}(\mu=2)$ can be obtained
analytically (this is nothing but the standard linear least-variance 
estimation).
For $1<\mu<2$, $\vctr{f}(\vctrg{\kappa},\mu)$ is bounded.
Its derivative with respect to $\vctrg{\kappa}$,
$\pdrv{}{\vctrg{\kappa}}\vctr{f}(\vctrg{\kappa},\mu)$, is
given by (\ref{eq:newdef}) which is always positive definite.
It can be singular and diverge to $+\infty$ due to absolute value terms
behaving like
$\lim_{x\to 0}|x|^{\mu-2}=\infty$, as can be seen in equation \ref{eq:Ok}).
The  derivative of $\vctr{f}(\vctrg{\kappa},\mu)$
with respect to $\mu$ is:
\ba\label{eq:dfKmudmu}
  \pdrv{}{\mu}  \vctr{f}(\vctrg{\kappa},\mu) & = &
  \pdrv{^{2}}{\vctrg{\kappa}\partial \mu} {\hat B}_{ii}
       \nonumber \\
  &=& \mu \sum_{p=1}^{N}  \left[ -
\biggl( \log \big|\left(G\spf_{ip}- 
\sum_{m=1}^{N}K\spo_{im}G\spf_{mp}\right)\big|\biggl)
  \left(G\spf_{ip}- \sum_{m=1}^{N}K\spo_{im}G\spf_{mp}\right)\pL{\mu-1}
   G\spf_{jp}C\spf_{p} \nonumber \right. \\
  & & \qquad +  \left.
  \biggl(\log \big|\left(\sum_{m=1}^{N}K\spo_{im}G\spo_{mp}\right) \big|\biggl)
   \left(\sum_{m=1}^{N}K\spo_{im}G\spo_{mp}\right)\pL{\mu-1}
   G\spo_{jp}C\spo_{p}\right]~~.
\ea
This derivative can also be singular due to the absolute value terms
behaving like
$\lim_{x\to 0} x\pL{\mu-1} \log|x| =\pm \infty$.
Note that $\pdrv{}{\vctrg{\kappa}}\vctr{f}(\vctrg{\kappa},\mu)$
and $\pdrv{}{\mu}  \vctr{f}(\vctrg{\kappa},\mu)$ become singular
simultaneously, but the former is more singular than the latter
because $\lim_{x\to 0} |x|^{\mu-2} / x\pL{\mu-1} \log|x| = \pm \infty$.
Implicit function theory can therefore be applied to guarantee that
a unique solution branch $\vctrg{\kappa}(\mu)$ exists
for $1<\mu<2$, starting from the analytical solution at $\mu=2$.
Indeed, if there exist other solution branches, then there must be at least one
bifuration as $\mu$ varies because the solution at $\mu=2$ is 
globally unique due to
linearity. The fact that $\pdrv{}{\vctrg{\kappa}}\vctr{f}$ is a 
positive definite matrix
globally (though it can be singular) however guarantees that there is no
bifuration.   Accordingly, the solution branch from $\mu=2$ provides the unique
solution of the system as $\mu$ varies.

The solution on the branch can be obtained numerically, either by
directly solving for the $N$ nonlinear equations for
each $\mu$ or by following the branch using the pseudo arc-length
continuation method \cite{arc-length}.

\newpage

\newpage

\newpage

\newpage

\newpage

\newpage

\newpage

\newpage

\newpage

\newpage

\newpage

\end{document}